\documentclass[conference,compsoc]{IEEEtran}
\IEEEoverridecommandlockouts

\usepackage[linesnumbered,algoruled,boxed,lined]{algorithm2e}
\usepackage{algpseudocode}
\usepackage{makecell}
\usepackage{amsmath}
\usepackage[numbers]{natbib}
\usepackage{multirow}
\usepackage{booktabs}
\usepackage{hyperref}
\usepackage{diagbox}
\usepackage{amssymb}
\hypersetup{
    colorlinks=true,
    linkcolor=blue,
    filecolor=magenta,
    urlcolor=magenta,
}
\usepackage[all]{hypcap}
\usepackage{booktabs}
\usepackage{url}
\usepackage{xcolor}
\usepackage{tikz}
\usepackage{diagbox}
\usepackage{listings}
\usepackage{comment} 
\usepackage{graphicx}
\usepackage[framemethod=TikZ]{mdframed}
\usepackage{tcolorbox}
\usepackage{amsthm}
\usepackage{colortbl}
\usepackage{caption}

\theoremstyle{definition}

\tcbset{
    colback=gray!10, 
    colframe=gray, 
    coltext=black, 
    boxsep=5pt, 
    arc=0mm, 
    top=0mm, bottom=0mm, 
    left=1mm, right=1mm, 
    boxrule=0.5mm, 
    toptitle=1mm, bottomtitle=1mm, 
    titlerule=0mm, 
}
\definecolor{deepgreen}{rgb}{0.0, 0.4, 0.0}

\definecolor{codegreen}{rgb}{0,0.6,0}
\definecolor{codegray}{rgb}{0.5,0.5,0.5}
\definecolor{codepurple}{rgb}{0.58,0,0.82}
\definecolor{backcolour}{rgb}{0.95,0.95,0.92}

\lstdefinestyle{mystyle}{
  backgroundcolor=\color{backcolour}, commentstyle=\color{codegreen},
  keywordstyle=\color{magenta},
  numberstyle=\tiny\color{codegray},
  stringstyle=\color{codepurple},
  basicstyle=\ttfamily\footnotesize,
  breakatwhitespace=false,         
  breaklines=true,                 
  captionpos=b,                    
  keepspaces=true,                 
  numbers=left,                    
  numbersep=5pt,                  
  showspaces=false,                
  showstringspaces=false,
  showtabs=false,                  
  tabsize=2
}
\usepackage{xurl}

\definecolor{mycolor}{RGB}{194, 214, 236}

\newcounter{finding}

\newcommand{\find}[1]{
\begin{tcolorbox}[leftrule=0.5mm,toprule=0mm,bottomrule=0mm,left=0.7pt,right=0.7pt,top=0.2pt,bottom=0.2pt]
\em #1
\end{tcolorbox}
}

\newcounter{result}

\makeatletter
\g@addto@macro{\@algocf@init}{\SetKwInOut{Parameter}{Parameters}}
\makeatother

\newcommand*\circled[1]{\tikz[baseline=(char.base)]{
            \node[shape=circle, draw, inner sep=0.2pt] (char) {\textcolor{black}{#1}};}}

\newcommand{\tech}{\mbox{\textsc{FirmHive}}}   
  
\begin{document}

\date{}

\title{LLMs as Firmware Experts:
A Runtime‑Grown Tree‑of‑Agents Framework}
 
\author{
\IEEEauthorblockN{Xiangrui Zhang\IEEEauthorrefmark{1}\thanks{Equal contribution.},
Zeyu Chen\IEEEauthorrefmark{2}\footnotemark[1],
Haining Wang\IEEEauthorrefmark{3},
Qiang Li\IEEEauthorrefmark{1}}
\IEEEauthorblockA{\IEEEauthorrefmark{1}Beijing Jiaotong University,
\IEEEauthorrefmark{2}Trinity University,
\IEEEauthorrefmark{3}Virginia Tech}\\
{\texttt{\url{https://github.com/bjtu-SecurityLab/firmhive}}}
}
 
\maketitle

\begin{abstract}

Large Language Models (LLMs) and their agent systems have recently demonstrated remarkable potential in automating code reasoning and vulnerability detection.
However, when applied to large-scale firmware, they perform poorly due to the firmware's binary nature, complex dependencies, and heterogeneous components. To address this problem,
this paper presents {\tech}, a recursive agent hive that enables LLMs to act as autonomous firmware security analysts.
{\tech} introduces two key mechanisms: (1) transforming delegation into a per‑agent, executable primitive and (2) constructing a runtime Tree of Agents (ToA) for decentralized coordination.
We evaluate {\tech} using real-world firmware images obtained from publicly available datasets, covering five representative security analysis tasks.
Compared with existing LLM‑agent baselines, {\tech} performs deeper (about 16$\times$ more reasoning steps) and broader (about 2.3$\times$ more files inspected) cross‑file exploration, resulting in about 5.6$\times$ more alerts per firmware.
Compared to state-of-the-art (SOTA) security tools, {\tech} identifies about 1.5$\times$ more vulnerabilities (1,802) and achieves 71\% precision, a significant improvement in both yield and fidelity.

\end{abstract}


\section{Introduction}

Large language models (LLMs) have demonstrated advanced capabilities across different domains such as knowledge acquisition, software development, game simulation, multi-robot systems, and adaptation to new scenarios~\cite{openai2024gpt4technicalreport, park2023generative, wu2024copilot, yang2024swe}. 
In the field of cybersecurity, LLMs have been increasingly used to automate tasks such as vulnerability scanning, malware detection, and the enforcement of security protocols~\cite{ullah2023llms, pearce2023examining, li2023hitchhiker, shang2024far, pearce2022pop, wang2024repository}. Traditionally, users have acted as intermediaries between models and specific tasks, manually interpreting model outputs and constraining their reasoning process. Recent advances in agentic LLMs have updated this paradigm by enabling autonomous systems that can perceive their environments, reason about goals, and execute tasks, using LLMs as their central decision-making component~\cite{yang2023intercodestandardizingbenchmarkinginteractive, wu2024oscopilotgeneralistcomputeragents, yang2024sweagentagentcomputerinterfacesenable, xie2024osworldbenchmarkingmultimodalagents}. This evolution allows agents to perform adaptive, goal-driven problem-solving with minimal human intervention. Beyond single-agent designs, multi-agent frameworks~\cite{hong2023metagpt, chan2023chateval} have emerged as scalable architectures that coordinate multiple specialized agents to enable distributed reasoning, collaborative decision-making, and complex system analysis across large-scale domains.

Firmware analysis presents a uniquely complex reasoning problem for LLM-based autonomous systems.
A single firmware image integrates binaries, shell scripts, and configuration files distributed across decentralized directory hierarchies.
Together, these properties introduce three technical challenges.
First, deep-chain reasoning: tracing a single vulnerability often requires following a long logical path across multiple files, from a top-level script to a specific control-flow segment within a binary.
Second, large-scale exploration: effective analysis must systematically traverse extensive filesystems to expose diverse attack surfaces.
Third, cross-component interaction reasoning: a vulnerability arises due to complex interactions among components, such as a misconfigured policy in a configuration file combined with an unsafe authentication routine. 
Existing LLM-based agent systems struggle to address these challenges of firmware analysis, making firmware a yet underexplored domain for LLM-based agent systems. 

In this paper, we present {\tech}, a firmware-oriented reasoning framework that enables LLMs to operate as autonomous firmware analysts.
{\tech} introduces two key innovations.
(1) Delegation as an ability transforms task delegation from a predefined workflow into an intrinsic agent capability, allowing any agent to dynamically decompose complex tasks, spawn sub-agents, and manage their results for both parallel and sequential reasoning.
(2) Tree of Agents (ToA) organizes these delegations into a runtime-evolving hierarchy whose depth and breadth adapt to task complexity.
Such a design partitions long reasoning chains into shorter, tractable steps, supports large-scale parallel exploration.
Together, these mechanisms enable {\tech} to maintain long-horizon reasoning, scale exploration, and integrate cross-file evidence—surpassing existing LLM-based agent systems.


We implement {\tech} through two cooperating modules: the Recursive Delegation Engine (RDE) and the Proactive Knowledge Hub (PKH).
The RDE module dynamically generates and coordinates agents and subtasks based on the firmware structure.
It decomposes long, monolithic reasoning chains into structured and parallel workflows, mitigating context fragmentation and enabling large-scale analysis across hundreds of interlinked components.
PKH acts as a persistent global memory that preserves semantic continuity across agents.
It continuously aggregates and reconciles intermediate results from multiple agents, recovering cross-component relationships and preventing redundant exploration.  
Together, RDE and PKH realize tree-guided yet graph-anchored reasoning: RDE aligns the reasoning process with the firmware hierarchy, while PKH maintains global coherence across distributed analyses.

To comprehensively assess {\tech}'s reasoning capabilities, we design five representative firmware analysis tasks (\S~\ref{sec:sub:setting}) that exhibit increasing levels of reasoning complexity.
The first two tasks (T1–T2) evaluate {\tech}'s ability to perform large-scale exploration and long-term reasoning by locating hard-coded credentials and identifying third-party component version information.
The next two tasks (T3–T4) extend this to deep analysis, requiring the system to trace the data flow of configuration variables and external inputs.
The final task (T5) represents full-stack vulnerability detection, integrating dependency tracing, dataflow reasoning, and contextual validation.

We evaluate {\tech} using the widely adopted Karonte firmware dataset~\cite{redini2020karonte}, which provides realistic and diverse firmware samples.
This choice ensures comparability with state-of-the-art (SOTA) systems, including Mango~\cite{gibbs2024operation} and SaTC~\cite{chen2021sharing}, while allowing us to stress-test multi-agent reasoning under complex firmware structures.  
Compared with existing LLM-agent baselines (Monolithic, MAS, MAS + Orchestrator), {\tech} demonstrates substantial improvements in reasoning depth and coverage, performing approximately 16× more reasoning steps and inspecting 2.3× more files, resulting in over five times more alerts per firmware.
Compared with SOTA tools, {\tech} autonomously identifies 1,802 vulnerabilities with 71\% precision, approaching the analytical capability of experienced firmware security experts.


The main contributions of this work are summarized as follows.

\noindent $\bullet$ We propose {\tech}, a firmware-oriented reasoning framework that organizes analysis through a dynamically evolving ToA.  
Its RDE adaptively decomposes complex reasoning tasks into localized subtasks aligned with firmware structure, and its PKH preserves long-term coherence and dependency integrity across distributed agents.

\noindent $\bullet$
We introduce two key mechanisms: (1) delegation as an ability, which allows every agent to spawn and manage sub-agents autonomously, and (2) a ToA that dynamically grows at runtime to balance reasoning depth and exploration breadth during firmware analysis.

\noindent $\bullet$ We implement and open-source {\tech}, providing the first end-to-end autonomous LLM-based framework for firmware analysis that operates without handcrafted rules or static pipelines.

\noindent $\bullet$ We conduct extensive evaluations across real-world firmware images and multi-layer analysis tasks, demonstrating that {\tech} achieves scalable, context-aware, and cross-component vulnerability discovery with higher precision and reasoning coverage than SOTA baselines.

\begin{figure*}[!t]
    \centering
    \includegraphics[width=5.7in ]{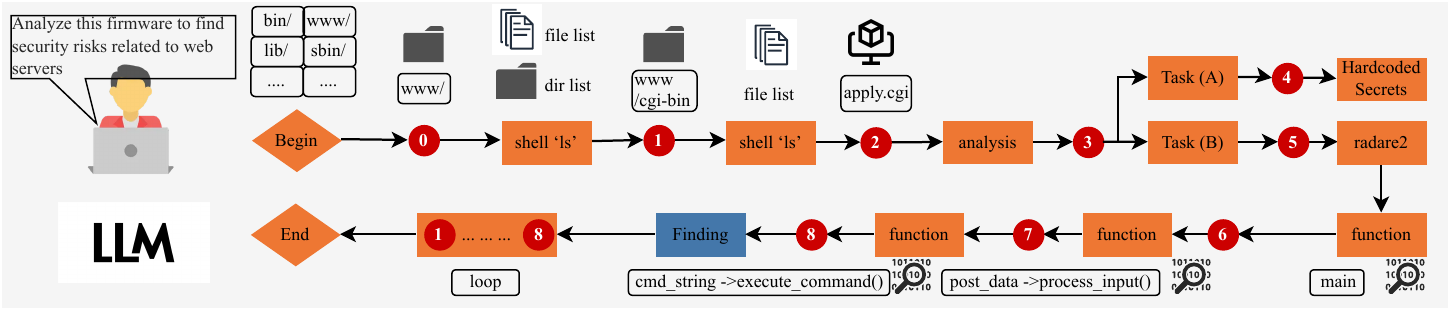}
    \caption{LLM as expert firmware analyst workflow. Red circles denote the LLM's reasoning turns; orange boxes denote tool invocations and their outputs.}
    \label{fig:expert:workflow}
\end{figure*}

\section{Motivation and Background}
\label{sec:back}

\subsection{Inherent Complexity of Firmware}

Firmware is the built-in software that initializes and controls hardware, manages drivers and system configuration, and provides essential runtime services.
It is deployed in routers, cameras, gateways, and a wide range of Internet of Things (IoT) devices.
Unlike traditional software codebases, firmware exhibits two defining characteristics.
(1) Vast and opaque structure.
A single firmware image may contain hundreds or thousands of files—binaries, configuration data, scripts, and web assets—interwoven into complex directory hierarchies.
This heterogeneity makes it difficult to locate and interpret security-relevant components.
(2) Intricate and implicit dependencies.
Firmware components are tightly coupled: scripts invoke binaries, configurations influence runtime behaviors, and startup routines coordinate entire subsystems.
These relationships are often implicit, dispersed across multiple files, and lack explicit documentation.
As a result, effective firmware analysis requires long-term tracing, large-scale exploration, and cross-file reasoning to uncover latent vulnerabilities.


\subsection{LLMs as Firmware Experts}

If an LLM is positioned as an autonomous firmware analyst, what systematic workflow would it follow?
To demonstrate this, we task the model with “{Analyze firmware (Archer\_C2\_V1\_170228) for web-related vulnerabilities.}”
Figure~\ref{fig:expert:workflow} illustrates this workflow, highlighting the extensive reasoning required across directories, files, functions, and data dependencies.

\textit{Step \circled{1}. Hierarchical exploration.}
The LLM enumerates top-level directories (e.g., via \texttt{ls}), identifies potential service folders such as /www, and progressively drills down—first to /www/cgi\_bin and then to candidate files like \texttt{apply.cgi}.

\textit{Step \circled{2}. File analysis.}
The LLM performs lightweight static inspection to determine file type (script or binary), extract symbols, search for hardcoded secrets, and locate external input points (e.g., HTTP POST, environment variables, or file reads).
It then selects appropriate toolchains—such as static taint analyzers or disassemblers—for deeper investigation.

\textit{Steps \circled{3}–\circled{5}. Deep reasoning.}
The LLM conducts parallel subtasks:
(A) searching for hardcoded credentials and
(B) performing taint propagation analysis.
For (A), the LLM scans for hardcoded credentials and configuration artifacts.
For (B), the LLM involves taint analysis: tracing external input propagation at function granularity while recording intermediate reasoning results, starting from the function \texttt{main}.

\textit{Steps \circled{6}–\circled{8}. Taint propagation Tracing.}
The model reasons over function-level granularity—tracking data flow from input buffers to sensitive sinks (\texttt{system()} and \texttt{execvp()}) while documenting call-stack traces, sanitization checks, and path constraints.
Each sub-analysis produces structured evidence (file path, code snippet, or call chain), which the LLM later aggregates into a vulnerability report.

\textit{Step \circled{n}. Iteration and aggregation.}
The process \{\circled{1}$\rightarrow\dots\rightarrow$\circled{8}\} repeats recursively for other directories and files until coverage is complete, producing consolidated findings—mirroring the systematic workflow of a domain expert.


\textbf{Key distinctions.}
Compared to conventional firmware analyzers~\cite{chen2021sharing, gibbs2024operation, redini2020karonte}, LLM-based experts exhibit two fundamental differences:
(1) \textit{Dynamic incorporation of intermediate artifacts.}
The LLM can read and incorporate many intermediate artifacts (e.g., temporary variables, file contents, and tool outputs). 
This is the primary distinction from traditional program‑analysis tools: by consuming intermediate evidence, an LLM can adaptively revise its strategy and analysis path, whereas conventional tools typically follow a fixed, pre‑defined pipeline.
(2) \textit{Leaving all firmware analysis jobs to LLMs.}
We delegate firmware analysis tasks to LLMs rather than encoding firmware‑specific logic as hardcoded scripts. As a result, our system can execute multiple tasks (T1–T5 in \S~\ref{sec:sub:setting}) without any modifications, whereas many traditional approaches are designed for a single, narrowly defined task.

\subsection{Limitations of LLM-based Agents}

To work effectively as expert firmware analysts, LLMs must operate for extended periods, often requiring many turns of reasoning and exploration.
Each turn involves tool invocations, output parsing, and thoughtful decision-making about next steps.
Current LLM-based agent architectures struggle to maintain such long-term operations.

\textbf{Monolithic Agent} 
A single agent, even with the ReAct pattern~\cite{yao2022react, wei2022chain,shinn2023reflexion}, is ill-equipped to handle long-term and large-scale reasoning. 
As shown in Figure~\ref{fig:expert:workflow}, tracing a vulnerability requires preserving a chain of facts across many steps (e.g., directory → subdirectory → config entry → CGI script → function → proportionated path).
We observe that a single agent inevitably forgets earlier, potentially crucial facts (e.g., the original source of tainted input), breaking the chain of evidence and degrading analysis accuracy.
The underlying cause is the limited context window and the stateless, single‑call nature of LLMs.

\textbf{Multiple Agent System} (without coordination) is the simplest form that expands analysis coverage by running multiple independent agents in parallel.  
Each agent explores its assigned roles within its own context and may use different tools or prompt templates.
However, multi‑agent coordination challenges are, in many respects, similar to those of multi-threaded concurrency: state aggregation, consistency, interference, task allocation, and communication~\cite{wu2023autogen, wang2024survey}.
Without coordination, MAS only gains breadth (agents operate in isolation), and depth reasoning chains are often lost.
In addition, MAS is statically configured: agent count and roles are fixed in advance, making it brittle for different firmware images and tasks.


\textbf{MAS with Orchestrator} (centralized coordination) is a star topology variant of MAS where an orchestrator agent manages and coordinates multiple subagents~\cite{wu2023autogen,chen2024autoagents,fourney2024magentic,wang2024survey}.
The orchestrator dynamically spawns subagents and aggregates subagent outputs for consumption by the lead agent. 
However, in firmware analysis, it is often impossible to anticipate the specifics of a given task, such as the number of files that need to be analyzed, the complexity of the call chains that must be traced, or the depth of the reasoning process that must be pursued. 
Thus, these complexities concentrate in the orchestrator: it inherits limited context windows and the stateless, single‑call behaviors of backend LLMs, becoming a capacity bottleneck for long-term and large-scale analyses.

\section{System Overview} 
\label{sec:design}

We propose {\tech}, where the LLM works as a human expert: it autonomously explores firmware to locate candidate files and performs detailed, step‑by‑step forensic analyses.
Unlike conventional MAS, in which orchestrators are in charge of task routing, subagent spawning, and evidence aggregation, {\tech} eliminates the use of orchestrators.
Task decomposition and subagent spawning are handled locally via delegation, treated as a per-agent capability (\S~\ref{sec:sub:ability}), while subtask routing, evidence flow, and result aggregation occur within the runtime-grown Tree of Agents (ToA), allowing distributed, context-aware coordination (\S~\ref{sec:sub:toa}).


\subsection{Delegation as an Agent Capability}
\label{sec:sub:ability}

Prior work treats delegation as a hardcoded orchestrator workflow, where only the orchestrator spawns sub-agents, generates subtasks, and aggregates results.
We propose a novel concept: delegation is a built-in capability available to every agent, enabling any agent to decompose tasks, spawn sub-agents, and manage subtasks based on its local context and workload.

\textbf{Delegation} refers to the process by which an agent decomposes a complex task into manageable subtasks, creates sub-agents, and assigns each subtask to the appropriate sub-agent for execution.
Making delegation an agent-level capability brings 3 benefits.

\begin{figure}[!t]
    \centering
    \includegraphics[width=3.0in ]{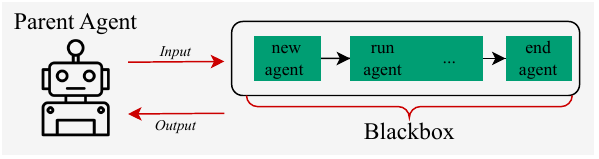}
    \caption{The delegation is viewed as a black box: the parent agent focuses on what needs to be delegated rather than how delegation is executed.}
    \label{fig:blackbox}
\end{figure}

\textit{(1) Delegation as a black box.} 
Figure~\ref{fig:blackbox} illustrates the black-box nature of delegation. 
A parent agent interacts minimally:
(i) construct a structured \texttt{subtask},
(ii) invoke \texttt{delegate(subtask)},
(iii) wait for execution, and
(iv) receive the result.
The internal process (instantiating sub‑agents, assigning subtasks, executing subtasks, and returning results) remain fully encapsulated.
This delegation keeps parent-side logic simple: it needs only to express the subtask and interpret the returned results, without understanding sub-agent internals.

\textit{(2) Context Isolation.} 
Delegation enforces strict context isolation between parent and sub-agents.
Each sub-agent runs in a fresh LLM instance with a clean context dedicated solely to its assigned subtask, preventing cross-interference and reasoning drift.
This isolation simplifies MAS coordination, and contributes to the stability of delegation—reflected in our stability results (\S\ref{sec:rq2}), where repeated runs yield highly similar agent counts and results.

\textit{(3) Recursive and decentralized delegation.}
Because delegation is a built-in capability rather than a hardcoded procedure, a dedicated orchestrator is no longer required.
Any agent can autonomously spawn sub-agents when needed, and each newly created agent inherits the same capability.
This recursive spawning gradually forms a dynamic and hierarchical agent topology that evolves at runtime according to task complexity and firmware structure.
Such decentralization eliminates the bottleneck of a central coordinator and lays the foundation for the runtime-grown ToA in the next subsection.

Delegation as an agent capability is essential: without it, workflows would remain static and unable to adapt to diverse firmware structures or tasks.
Existing frameworks (e.g., LangGraph~\cite{langgraph,langchain}, AutoGen, CrewAI) support delegation only through hardcoded control flows; they do not expose delegation as a first-class capability.
We therefore implement a custom framework where delegation operates as an agent-level ability rather than a fixed code path.

\begin{figure}[!t]
    \centering
    \includegraphics[width=2.5in ]{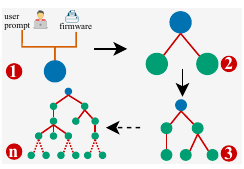}
    \caption{A runtime grown ToA: the root agent recursively spawns sub-agents via delegation and forms a tree structure.}
    \label{fig:tree}
\end{figure}

\subsection{Tree of Agents (ToA)}
\label{sec:sub:toa}

Building on the delegation capability, we propose the \textit{Tree-of-Agents (ToA)}, a dynamic hierarchical structure that leverages recursive agent delegation to organize firmware analysis tasks. 
Each agent can autonomously spawn sub-agents and decompose its assigned task, and these delegation actions collectively form the ToA.

\textbf{Definition.} \textit{Tree of Agents (ToA)} is a hierarchical structure of agents, where each node represents an agent and each edge represents a delegation from a parent agent to a child.

Figure~\ref{fig:tree} illustrates a runtime-grown ToA.
At the start, a single root agent receives the user prompt and firmware image. Through recursive delegation, this root agent incrementally expands the ToA, spawning sub-agents as needed to accomplish subtasks.

\textit{Dynamic Nature.}
Unlike a fixed, predesigned tree with predetermined nodes, roles, or depth, the ToA evolves dynamically at runtime.
Each node autonomously decides whether to delegate subtasks to child nodes based on its local context and task progress. 
Consequently, the overall tree structure—including the number of nodes, branching factor, and depth—is adaptively determined by task complexity and reasoning demands.
This dynamic growth is crucial for firmware analysis: firmware images differ in layout, file types, and file contents, and vary in user tasks.

\textbf{ToA Constraints.}
Since delegation is treated as an agent capability, each agent autonomously decides when and how to delegate based on its local context.
However, unconstrained delegation can cause uncontrolled spawning of sub-agents, redundant computation, and excessive resource consumption.
We therefore enforce several invariants to ensure that the ToA remains bounded and predictable.

\noindent
(1) \textit{Task \& Context Isolation.} 
    Each agent operates strictly within its (task, context) scope, with no shared mutable local state across agents.
    Task and context are determined by firmware structure, as detailed in \S~\ref{sec:sub:task}.

\noindent
(2) \textit{Parent–Child Delegation \& Communication Only}. 
    Agents may delegate tasks solely to their immediate children (\S~\ref{sec:sub:delegation}).
    Communication is restricted to parent–child pairs.
    Cross-branch delegation or arbitrary messaging is forbidden (\S~\ref{sec:sub:comm}).

\noindent
(3) \textit{Structural Evolution.} 
    The ToA cannot evolve arbitrarily at runtime. 
    Its growth strictly follows the inherent structure of firmware—limited to three agent types (\textit{directory}, \textit{file}, and \textit{function} in \S~\ref{sec:sub:type})—and is further constrained by recursion depth and branching limits (\S~\ref{sec:sub:blueprint}).

\noindent
(4) \textit{ToA Termination.} 
   Each agent has explicit termination conditions: successful task completion, budget exhaustion (step/time), or unrecoverable failure (including child or LLM errors).  
The entire ToA converges when the root agent terminates.

To compensate for the loss of cross‑agent messaging and to provide necessary global visibility, we introduce a \textit{Persistent Knowledge Hub (PKH)} (\S~\ref{sec:sub:pkh}). 
Agents publish their findings to the PKH, which serves as a durable, queryable shared memory enabling cross-branch correlation and high-level synthesis.

\section{{\tech}: System Design}
\label{sec:system}

\begin{figure}
    \centering
    \includegraphics[width=3.3in]{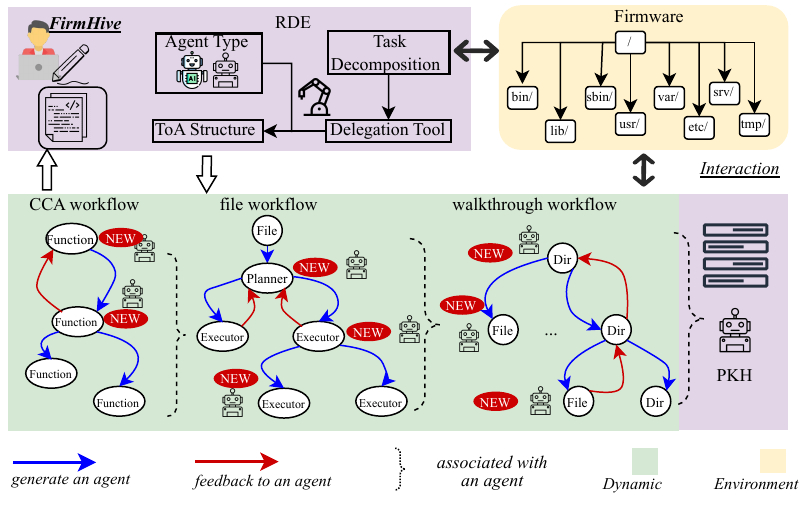}
    \caption{{\tech}'s architecture: the purple region represents the RDE and PKH components; the yellow region represents the firmware environment, and the green region represents the runtime growth ToA.}
    \label{fig:arch:new}
\end{figure}

This section describes the design of {\tech}, which builds the ToA at runtime from a single root agent. 
The root agent takes a firmware image \(f\) and a user prompt \(u\) as input and runs a recursive procedure \(\mathcal{R}\) that spawns and coordinates child agents:
\begin{equation} \label{eq:toa}
\mathrm{ToA} \;=\; \mathcal{R}\big(\mathrm{RootAgent}(f,u)\;\mid\;RDE,\;PKH\big)
\end{equation}
The root agent decomposes work based on what it observes in the firmware and on the current task context, and it creates subagents as needed.
The system does not require an explicit construction algorithm; instead, agent delegation under the Recursive Delegation Engine (RDE) and the Persistent Knowledge Hub (PKH) determines the ToA. Figure~\ref{fig:arch:new} shows the overall architecture. Sections~\ref{sec:sub:rde} and~\ref{sec:sub:pkh} describe RDE and PKH in detail.

\subsection{Recursive Delegation Engine (RDE)} 
\label{sec:sub:rde}

The RDE consists of four core components: task decomposition, delegation tool, agent types, and ToA structural stability.

\subsubsection{\textbf{Task Decomposition as Object-Target Reasoning}}
\label{sec:sub:task}

We adopt a novel decomposition paradigm based on an object-target formulation. 
Specifically, we treat each task as a pair: \textbf{(object, analysis target)}. 
The object refers to an element in the firmware—such as a file, directory, or function—while the target represents the analysis goal to be performed on that object.
We formalize this decomposition process as:
\begin{equation} \label{eq:decomposition}
Task(o, g) \xrightarrow{\text{observe}} \mathcal{O} \xrightarrow{\text{LLM}, \mathcal{C}} \{T_{sub}(o_i, g_i)\}_{i=1}^{k}
\end{equation}
When an agent has a \( Task(o, g) \), it observes the environment from its object \( o \), and extracts all observable objects \( \mathcal{O} \).
Then, the agent will pick up relevant objects \( o_i \in \mathcal{O} \) and reason about the analysis targets \( g_i \) for each selected object.
Thus, the original task is decomposed into \( k \) subtasks \( \{T_{sub}(o_i, g_i)\}_{i=1}^{k} \).

\textit{Current Context} 
\( \mathcal{C} \) encapsulates 2 key elements:
(i) the user prompt, which defines the global analysis goal; and (ii) the agent's history, including prior observations and tool outputs, which provide factual grounding accumulated during recursive decomposition.

\textit{Observation} $\mathcal{O}$ denotes the set of observable objects that an agent extracts from its assigned object $o$ (for example, a directory listing, file contents, or a function's disassembly).
$\mathcal{O}$ is determined by the firmware image and the runtime inspection procedure. 
Hence, for a fixed firmware and object $o$, the resulting observation $\mathcal{O}$ is effectively immutable during the decomposition step.
This stability bounds the agent's perceptual nondeterminism and confines LLM reasoning to a well-scoped local view.

\textit{Controlled Autonomy in Decomposition.}
Although the decomposition itself is left to the LLM, it is not arbitrary.
Each decision is bounded by two stable anchors—(a) the context and the (b) observation—both determined by the firmware.
Formally, the decomposition process can be viewed as:
\[
\underbrace{(\text{Context},\ \text{Observation},\ \text{Prompt})}_{\text{fixed}} 
\xrightarrow{\text{LLM (variable)}} 
\text{Delegation}
\]
Within this decision chain, only the LLM’s reasoning introduces variability. 
By constraining the other components, the delegation behavior remains both controllable and reproducible.
In practice, we further reduce nondeterminism (e.g., setting temperature=0 and fixing random seeds), ensuring that recursive decompositions converge to a structure-consistent ToA.

\subsubsection{\textbf{Delegation Tool}}
\label{sec:sub:delegation}

Delegation represents a fundamental capability of an agent—its ability to distribute tasks to other agents. 
{\tech} encapsulates this capability into a reusable \textit{delegation tool}, transforming what was traditionally handcrafted control logic into a unified and composable runtime abstraction. 
Any agent that installs this tool automatically acquires the ability to spawn, manage, and collect results from sub-agents. 
Thus, delegation becomes an agent-level capability rather than a hardcoded function.

\textit{Agent Initialization.}
Each agent is instantiated through a reusable configuration abstraction:
{\small
\begin{equation}\label{eq:agent-config}
\textit{AgentConfig = (llm, output\_schema, memory, task, type, tool)}
\end{equation}}
where $llm$ denotes the reasoning model; $output\_schema$ constrains structured responses; $memory$ preserves contextual state; $task$ binds the concrete object–target pair $(o,g)$; $type$ specifies the agent’s scope (e.g., directory, file, function); and $tool$ enumerates its available tools. 
This uniform interface allows newly spawned agents to be created dynamically and consistently across the system.

\textit{Execution and Result Return.}
The execution of a delegated sub-agent is equivalent to invoking a tool call: a parent agent delegates a task, the sub-agent performs reasoning and analysis, and returns structured results to its parent.
This return mechanism forms the basis of parent–child communication (\S~\ref{sec:sub:comm}).

\begin{figure}[!t]
    \centering
    \includegraphics[width=2.3in ]{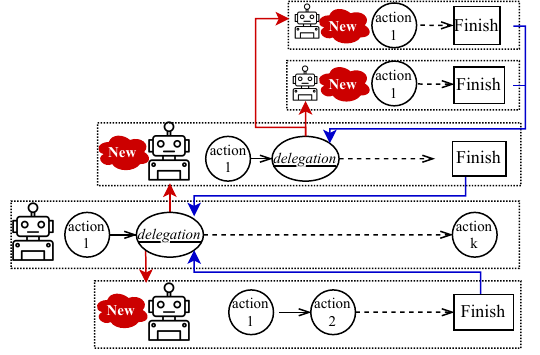}
    \caption{Recursive delegation constructs a tree: the red arrow represents the delegation, and the blue arrow indicates the return of results. }
    \label{fig:delegation}
\end{figure}

\textit{Delegation Patterns.}
The tool supports two canonical delegation modes (Table~\ref{tab:tool} in Appendix):  

\noindent$\bullet$ \textit{Parallel Delegation for Breadth (1$\rightarrow$n)} — An agent handling large-scale workloads (e.g., multiple binaries or configuration files) can spawn multiple sub-agents in parallel to accelerate coverage.  

\noindent$\bullet$ \textit{Chained Delegation for Depth (1$\rightarrow$1)} — For sequential reasoning tasks (e.g., taint tracking along a function call chain), an agent delegates tasks step by step, extending the reasoning path vertically.

\textit{Recursive Delegation.}
Each spawned agent inherits the same delegation tool, enabling recursive task propagation. 
As shown in Figure~\ref{fig:delegation}, the resulting delegation hierarchy forms a dynamic tree: a parent agent (A) delegates subtasks to B1 and B2; B1 further delegates to C1 and C2.
The parent agent awaits feedback from the subagents: A waits for results from B1 and B2; B1 waits for results from C1 and C2.
If the parent performs approximately 3 reasoning steps and each child performs 4, recursive delegation extends the reasoning capacity to roughly $3\times4\times4=48$ sequential steps (plus integration).
This recursive mechanism allows both scalable breadth (parallel exploration) and extended depth (sequential reasoning) in complex firmware environments.

\subsubsection{\textbf{Agent Types}}
\label{sec:sub:type}

{\tech} defines three agent types, each operating under the unified ReAct loop~\cite{yao2022react,schick2023toolformer}:
\[
\{\text{observe} \rightarrow \text{reason} \rightarrow \text{act} \rightarrow \text{observe} \dots \}
\]
They differ in operational scope, reasoning granularity, and the system prompts that define their reasoning context, yet all share a unified delegation mechanism and evidence-grounding constraint.
The exact system prompts are provided in Appendix~\ref{prompt}.

\textbf{Directory Agent} explores the firmware filesystem at directory granularity. 
Given a task $T(o, g)$, where $o$ is the directory object and $g$ the analysis goal, the agent scans directory contents, enumerates subdirectories and files, and extracts metadata. 
When deeper inspection is needed, it selects candidate objects $o_i$ and infers corresponding subgoals $g_i$, delegating them to sub-agents. 
Its mission is exploratory—identifying suspicious objects for downstream analysis.
The agent terminates once all potential targets are analyzed or when explicitly halted by its parent, returning structured evidence to the upper layer.

\textbf{File Agent} performs targeted, evidence-driven analysis on a single file object. 
Given $T(o, g)$, it inspects the file according to the goal derived from the user prompt. 
Lightweight files (e.g., configuration or script) are processed locally, whereas binaries requiring deeper inspection are delegated to function agents. 
Like directory agents, file agents operate under the ReAct cycle until all subtasks complete or upon receiving a stop signal.

\textit{Evidence-first policy.}
Each finding must be backed by tool outputs and include provenance metadata (agent\_id, tool, timestamp, raw evidence snippet). 
If no evidence exists, the file agent reports uncertainty or requests validation, preventing unsupported hallucinations.

\textbf{Function Agent} performs call-chain analysis (CCA) within a binary function.
Given a task $T(o, g)$, where $o$ includes the current callee and its caller context, it traces taint propagation within the function’s scope in a forward-only manner.
If propagation extends into new callees, the agent selects them as new objects $o_i$ and infers their goals $g_i$, delegating them to newly spawned sub-agents.
Execution continues until all delegated subtasks return or the agent is explicitly stopped.

\textit{Evidence-grounded tracing.}
Each propagation path must be supported by tool evidence (e.g., r2 analysis output) and accompanied by trace metadata (trace type, identifier, address, code snippet).
If a trace is interrupted, the agent must explicitly record the stop point and reason, ensuring reproducibility and verifiability of all findings.



\subsubsection{\textbf{ToA Structural Growth}}
\label{sec:sub:blueprint}

Building on the general design principles introduced earlier, this subsection elaborates on how the ToA grows and self-organizes during analysis.
The growth process follows a hierarchical blueprint that specifies, for each object type:
(i) which agent type should be instantiated,
(ii) which target objects the agent is responsible for, and
(iii) the permissible delegation patterns and runtime constraints.

\textbf{Directory Analysis Layer.} 
This layer serves as the entry point for firmware filesystem exploration.
Directory agents enumerate directory entries and perform two types of delegation:
(a) spawning new directory agents for subdirectories ($\langle\text{dir}\rightarrow\text{subdir}\rangle$), and
(b) spawning file agents for file objects ($\langle\text{dir}\rightarrow\text{file}\rangle$).
This recursive but bounded delegation model enables adaptive and scalable exploration aligned with firmware’s hierarchical nature.

\textbf{File Analysis Layer.}
At this layer, a file agent conducts a focused analysis tailored to the file type. 
A file agent applies two main delegation patterns:
(a) decomposing complex files into concurrent file-level subtasks ($\langle\text{file}\rightarrow\text{file}\rangle$) to enable parallel checks, and
(b) for compiled binaries, spawning function agents for deeper inspection ($\langle\text{file}\rightarrow\text{function}\rangle$).
This decomposition emphasizes parallelism over recursion, allowing efficient task expansion without uncontrolled growth.

\textbf{Call-Chain Analysis Layer.}
At this layer, function agents perform call‑chain analysis for binaries. 
Each function agent first identifies source functions, which serve as entry points in the function call graph.
(a) For each source function, the agent spawns tracer agents ($\langle\text{function}\rightarrow\text{function}\rangle$) to follow call paths toward sensitive operations (e.g., \texttt{system}, \texttt{strcpy}, \texttt{sprintf}).
(b) Long paths are handled via chained delegation (sequential subagents) to extend reasoning depth while respecting predefined depth and reasoning-step limits.
The workflow mirrors the Planner → Executor pattern: the function agent plans the call-chain traversal, and tracer agents execute the stepwise propagation tracing. 
Multiple source functions are handled in parallel to ensure breadth of coverage.

\textbf{Growth Constraints.}
{\tech} enforces several structural constraints to maintain tractable ToA growth and stable reasoning.
First, \textit{task isolation} guarantees non-overlapping task scopes—each subtask is anchored to a concrete object, ensuring agents never analyze the same target concurrently.
Second, recursive delegation in the Directory and Call-Chain layers is bounded by a maximum depth $D_{\max}$, while the File layer enforces non-recursive parallel decomposition.
Finally, global execution limits—such as per-agent reasoning-step bound $S_{\max}$.
Together, these constraints align ToA expansion with firmware structure while preserving stability and determinism.

\subsection{Persistent Knowledge Hub (PKH)}
\label{sec:sub:pkh}

\begin{figure}[!t]
    \centering
    \includegraphics[width=3.0in]{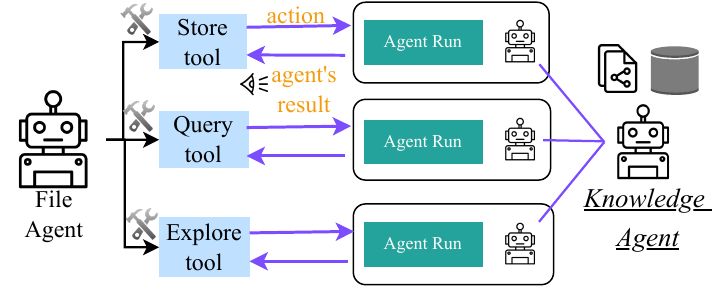}
    \caption{PKH architecture: each tool invocation is equal to the agent's execution; this agent is a global agent with a persistent store.}
    \label{fig:blueprint:pkh}
\end{figure}

While the ToA ensures distributed reasoning and structured delegation, it lacks a global memory for persistent knowledge sharing and cross-branch synthesis. 
To address this limitation, we introduce the PKH as shown in Figure~\ref{fig:blueprint:pkh}.
Similar to the delegation tool, PKH is encapsulated as a reusable agent-like tool. 
The key difference is that PKH operates as a \textit{global agent} with persistent storage—it is not instantiated per task. 
It maintains a structured repository of findings across all agents and branches within the ToA, serving as a long-term, queryable knowledge base.


PKH provides a controlled shared store to prevent information loss and enable safe cross-branch correlation and synthesis. 
Specifically, PKH exposes three tools to agents:
\begin{itemize}
\item \textbf{Store}: submit a new structured finding to PKH.
\item \textbf{Query}: retrieve relevant findings from PKH.
\item \textbf{Explore}: enumerate potential findings that may correlate with the current context.
\end{itemize}
When equipped with these tools, an agent can extend its local context into a global context by querying PKH. 
By correlating evidence across branches, agents can synthesize higher-level insights—for example, linking a library-local buffer overflow with an independent finding showing that the library handles privileged external input.

\textit{Structured schema.} 
To maintain consistency and enable machine reasoning, PKH organizes all stored findings into a structured schema capturing four essential dimensions:
(i) the finding itself,
(ii) its observed location,
(iii) its semantic type (e.g., vulnerability or misconfiguration), and
(iv) linking identifiers that serve as anchors for cross-record correlation.
This schema provides a uniform representation for agents to record, retrieve, and correlate evidence across different branches.

\textit{Proactive correlation loop.}  
Unlike RAG systems, PKH performs proactive reasoning over new findings through an \textit{explore → query → enrich → store} loop: 
(1) explore related findings, (2) query matching entries, (3) enrich and validate the current finding, and (4) store the final, enriched record. 
This loop distinguishes PKH from passive RAG systems: rather than merely retrieving context, PKH actively contextualizes new evidence as it is produced, enabling on-the-fly synthesis and high-level insight formation.

\textit{Concurrency and Consistency.}  
Multiple agents may access PKH concurrently, risking duplication or partial overwrites. 
To ensure safe concurrent access, PKH employs lightweight file-level locking (\texttt{fcntl.flock}) to serialize write operations, guaranteeing atomic updates. 
Only file agents are permitted to read from and write to PKH. 
Before committing a finding, a file agent must (i) query related entries, (ii) validate supporting evidence, and (iii) deduplicate subagent outputs. 
This disciplined access model preserves both concurrency safety and semantic integrity across the agent hierarchy.

\subsection{Inter‑Agent Communication}
\label{sec:sub:comm}

Inter-agent communication is intentionally constrained to avoid cross-scope interference and to maintain task execution.
The system permits only three channels: (i) Parent$\rightarrow$Child delegation, (ii) Child$\rightarrow$Parent result return, and (iii) File Agent$\leftrightarrow$PKH interactions.

\textit{Parent\(\rightarrow\)Child.} 
A parent agent passes a structured tuple \((o,g)\) (\S~\ref{sec:sub:task}) to the delegation tool (\S~\ref{sec:sub:delegation}).
No implicit state or shared memory is passed.

\textit{Child\(\rightarrow\)Parent.} 
A child agent returns a structured, LLM-summarized result that the parent consumes as a function-like return value. 
(1) A function agent reports function-level findings, including vulnerability classification (e.g., CWE), propagation path (source–sink trace), code or assembly snippets for each step, and a machine-parsable trace fragment (addresses, function identifiers, symbolic references).  
(2) A file agent returns aggregated file-level findings that summarize evidence across all internal analyses and subagents.
Its format is the same as PKH's structured schema (\S~\ref{sec:sub:pkh}).
(3) A directory agent reports all file-level findings aggregated from its children.

\textit{File Agent\(\leftrightarrow\)PKH}. 
Only file agents are authorized to communicate with the PKH.  
They use the same structured schema for findings and may invoke the \texttt{store}, \texttt{query}, or \texttt{explore} tools.  
Directory agents are excluded since they already aggregate child findings, and function agents are limited to local scopes to prevent global interference.

\textit{Evidence-First Rule.}  
All agent communications must be grounded in verifiable evidence.  
Each finding (file‑ or function‑level) must be grounded in tool outputs or raw artifacts. 
If an evidence piece is missing, the return must be marked \texttt{partial} with an explicit explanation and concrete follow‑ups.

\section{Evaluation}
\label{sec:exp}

This section demonstrates the effectiveness of {\tech} via real-world experiments.
Specifically, we aim to answer the following research questions:

\noindent $\bullet RQ1:$ How does {\tech} perform compared to state-of-the-art (SOTA) firmware analysis tools? (\S~\ref{sec:rq4})

\noindent $\bullet RQ2:$ How does {\tech} compare with existing LLM-based agent architectures? (\S~\ref{sec:rq1})

\noindent $\bullet RQ3:$ To what extent are {\tech}'s analysis results stable and reproducible? (\S~\ref{sec:rq2})

\noindent $\bullet RQ4:$ How do {\tech}'s key design components contribute to its overall performance? (\S~\ref{sec:rq3})

\subsection{Implementation \& Experimental Setup} 
\label{sec:sub:setting}

\textbf{Prototype System}.
We implemented a full prototype of {\tech} in Python.
The system integrates LLM-based agents, tool invocation interfaces, and a PKH.
{\tech} is implemented from scratch rather than relying on existing orchestration frameworks (e.g., LangChain, LangGraph), as firmware analysis requires fine-grained control over agent delegation and execution flow.
Detailed implementation and tool descriptions are provided in Appendix~\ref{appendix:imple}.

\textbf{Experimental Setup.}
All experiments are conducted with the backend LLM configured at a temperature of 0 and a fixed random seed to ensure deterministic outputs.
Each spawned agent operates with an independent LLM client instance, guaranteeing isolated context and preventing cross-agent memory leakage.
To bound runtime exploration, we set a per-agent reasoning limit of $S_{\max}=30$ steps and a maximum delegation depth of 4.
Each LLM query is retried up to three times upon failure; if all retries fail, the request is treated as a service failure and the corresponding subtask is recorded as \textit{partial} or \textit{deferred}.
These hyperparameters are selected empirically based on stability and responsiveness observed during pilot testing.

\textbf{Datasets.}
We evaluate {\tech} using the Karonte dataset~\cite{redini2020karonte}, which has also been adopted in Mango~\cite{gibbs2024operation} and SaTC~\cite{chen2021sharing}.
Each firmware contains numerous directories and files, underscoring the need for both large-scale exploration and long-horizon reasoning—tasks that static, non-adaptive agent workflows often fail to handle efficiently.
Detailed dataset statistics are provided in Appendix~\ref{appendix:toa_case} (Table~\ref{tab:dataset}).


\textbf{Baselines.}
We compare {\tech} against four baselines:

(1) \textit{State-of-the-Art (SOTA) Tools}:  
We use two widely recognized firmware vulnerability analysis tools, SaTC~\cite{chen2021sharing} and Mango~\cite{gibbs2024operation}.
These tools employ advanced static analyses, such as taint tracking, and represent highly optimized “expert knowledge” for firmware vulnerability discovery.

(2) \textit{SWE-Agent~\cite{yang2024sweagent} (with and without Knowledge Base)}:
SWE-Agent is a monolithic, React-style LLM agent tailored for software engineering tasks.
We configure SWE-Agent to use the same toolset and prompts as {\tech} to ensure a fair comparison.
We further evaluate SWE-Agent with a KB that allows long-term storage and retrieval of intermediate findings to support extended reasoning.

(3) \textit{Multi-Agent System (MAS) with Static Pipeline}:
To our knowledge, no off-the-shelf MAS is designed specifically for firmware analysis.
We implement a static MAS using AutoGen~\cite{wu2023autogen}.
First, MAS enumerates all top-level directories and initiates a corresponding number of directory agents. 
Second, MAS enumerates all files in each directory, in parallel, and launches a file agent for each file discovered. 
Finally, MAS further initiates function agents for each binary file  for deep analysis.
All agents send their findings back to the root agent for aggregation and reporting.

(4) \textit{MAS with Orchestrator}:
We reimplemented MAS from~\cite{chen2024autoagents,fourney2024magentic}, where we adapted their architectures to better suit the firmware analysis.
In this model, an orchestrator agent plans tasks, tracks progress, generates subtasks, and delegates them to worker agents.
Only the orchestrator can perform delegation; worker agents solely execute assigned tasks and report results back to the orchestrator.

\textbf{Task}.
We provide a set of tasks for {\tech} to complete, which are designed to cover firmware security analysis. 
Each task requires expertise that typically only trained firmware security professionals possess.
Successful completion of these tasks by {\tech} demonstrates that LLM-based agents can approach, or even achieve, expert-level performance in firmware analysis.

\textit{T1: Find Hard-coded Credentials}. 
\vspace{1mm}
\begin{mdframed}[backgroundcolor=gray!6,innertopmargin=4pt,innerbottommargin=4pt,leftmargin=0pt,rightmargin=0pt]
  User Prompt: Report all hard-coded credentials and other sensitive information.
\end{mdframed}
T1 is important because hard-coded credentials can grant attackers direct access, making them a critical target for firmware security auditing.

\textit{T2: Identify Third-party Components}.
\vspace{1mm}
\begin{mdframed}[backgroundcolor=gray!6,innertopmargin=4pt,innerbottommargin=4pt,leftmargin=0pt,rightmargin=0pt]
  User Prompt: Identify third-party components and their versions, generating a Software Bill of Materials (SBOM).
\end{mdframed}
T2 matters as it uncovers outdated third-party components linked to known CVEs, allowing defenders to assess and mitigate inherited risks.

\textit{T3: Tracing NVRAM/Environment Variable Interactions}.
\vspace{1mm}
\begin{mdframed}[backgroundcolor=gray!6,innertopmargin=4pt,innerbottommargin=4pt,leftmargin=0pt,rightmargin=0pt]
  User Prompt: Analyzing how the firmware interacts with configuration systems like NVRAM or environment variables (e.g., via \texttt{getenv}) to identify unsanitized data flow paths from these variables to dangerous function calls.
\end{mdframed}
T3 plays a key role in revealing how environment variables influence system behavior, often exposing subtle misconfigurations or security flaws.

\textit{T4: Web Attack Chain Analysis}.
\vspace{1mm}
\begin{mdframed}[backgroundcolor=gray!6,innertopmargin=4pt,innerbottommargin=4pt,leftmargin=0pt,rightmargin=0pt]
  User Prompt: Identifying how untrusted input from external HTTP requests is processed and tracing data flows to find paths where this input reaches dangerous functions (e.g., \texttt{system}, \texttt{strcpy}).
\end{mdframed}
T4 is crucial for mapping how untrusted web input can trigger dangerous operations, helping to expose web-based attack vectors like command injection.

\textit{T5: Vulnerability Detection and Verification }. 
\vspace{1mm}
\begin{mdframed}[backgroundcolor=gray!6,innertopmargin=4pt,innerbottommargin=4pt,leftmargin=0pt,rightmargin=0pt]
  User Prompt: Discover and report complete, end-to-end exploit chains within the firmware, specifically focusing on viable data flow paths from any untrusted input source to a security-critical sink.
\end{mdframed}
T5 is essential for uncovering deeply embedded, unknown flaws in firmware binaries and configurations through in-depth vulnerability analysis.

\begin{table}[!t]
\small
\centering
\caption{Comparison of {\tech}'s performance against SOTA tools. }
\label{tab:sota_e2e_comparison}
\setlength{\tabcolsep}{4pt}
\begin{tabular}{lcccc} 
\toprule
 & \makecell[c]{Binary \\ (CI+BOF)} & \makecell[c]{Binary \\ (Others)} & \makecell[c]{Non-Binary \\ Alerts} & \makecell[c]{\textbf{Total}} \\
\midrule
\textsc{SaTC}\cite{chen2021sharing} & 644 & 0 & 0 & 644 \\ 
\textsc{Mango}\cite{gibbs2024operation} & 1,109 & 0 & 0 & 1,109 \\
\midrule
\textsc{{\tech}}       & \textbf{1,206} & \textbf{192} & \textbf{404} & \textbf{1,802} \\ %
\bottomrule
\end{tabular}
\end{table}

\begin{table}[!t]
\small
\centering
\caption{Manual Evaluation of {\tech}'s alerts in T5.}
\label{tab:ai_verification}
\setlength{\tabcolsep}{5pt}
\begin{tabular}{l|cc|c}
\toprule
Alert & \multicolumn{2}{c|}{Manual} & Num. \\
& Vulnerable & Not Vulnerable & ($n$) \\
\midrule
Classified as `TP' & \textbf{142} & 42 & 200 \\
Classified as `FP' & 26 & \textbf{74} & 100 \\
\bottomrule
\end{tabular}
\end{table}

\subsection{Comparison with SOTA Tools via T5 (RQ1)}
\label{sec:rq4}


To answer RQ1, we leverage Task T5 (vulnerability detection) to compare {\tech} with two SOTA tools: Mango and SaTC. 
Task T5 represents the most comprehensive type of firmware security analysis, requiring the identification and validation of complete exploit chains—from untrusted input sources to critical sinks.
If {\tech} achieves comparable or superior performance to these SOTA tools, it provides strong evidence that LLMs can function as practical firmware security experts.
All experiments in this comparison use DeepSeek-v3.1 as the backend LLM.

We use the metric “alert” to represent a detected taint-based vulnerability within the firmware.
An alert is considered a verified finding if {\tech} confirms a complete, unsanitized data flow from an untrusted source to a critical sink.
To assess the accuracy of these alerts, we manually examine each case to determine whether it corresponds to a real vulnerability.
For fairness, we adopt the same threat model as \textsc{Mango}~\cite{gibbs2024operation}: an alert is labeled as a true positive (TP) only if it represents a vulnerability exploitable by an attacker with legitimate device access (e.g., connected via Wi-Fi with valid login credentials).

\begin{table*}[t]
\centering
\caption{Performance comparison between {\tech} and baselines across five tasks (T1-T5).}
\setlength{\tabcolsep}{1pt}
\begin{tabular}{l|ccc|ccc|ccc|ccc|ccc}
\toprule
\multirow{2}{*}{} 
& \multicolumn{3}{c|}{\textbf{T1}} 
& \multicolumn{3}{c|}{\textbf{T2}} 
& \multicolumn{3}{c|}{\textbf{T3}} 
& \multicolumn{3}{c|}{\textbf{T4}} 
& \multicolumn{3}{c}{\textbf{T5}} \\
& \textit{Alerts} & \textit{Step} & \makecell{\textit{Files} \\ \textit{accessed}} 
& \textit{Alerts} & \textit{Step} & \makecell{\textit{Files} \\ \textit{accessed}} 
& \textit{Alerts} & \textit{Step} & \makecell{\textit{Files} \\ \textit{accessed}} 
& \textit{Alerts} & \textit{Step} & \makecell{\textit{Files} \\ \textit{accessed}} 
& \textit{Alerts} & \textit{Step} & \makecell{\textit{Files} \\ \textit{accessed}} \\
\midrule
SWE 
& 4.2 & 43.2  & 8.6 & 5.6 & 49.1 & 7.3 & 3.4 & 47.6 & 5.6 & 0.4 & 41.8 & 9.3 & 1.1 & 47.1 & 5.3 \\
SWE +KB
& 4.4 & 48.0 & 7.5 & 6.20 & 44.3 & 7.2 & 3.8 & 42.9 & 5.6 & 0.2 & 46.2 & 9.4 & 1.4 & 49.2 & 6.1 \\ 
MAS+pipeline
& 4.6 & 70.5 & 70.3 & 5.84 & 152.3 & 100.5 & 3.2 & 300.4 & 70.3 & 1.1 & 80.5  & 41.4 & 1.4 & 290.2  & 121.2 \\ 
MAS+orchestrator
& 4.0 & 124.5 & 18.6 & 9.7 & 173.9 & 24.9 & 3.9 & 108.9 & 17.4 & 1.1 & 102.9 & 15.2 & 2.1 & 430.3 & 19.4 \\ \hline
\textbf{{\tech}} 
& \textbf{10.3} & \textbf{692.5} & \textbf{57.2} & \textbf{28.5} & \textbf{1718.2} & \textbf{96.5} & \textbf{12.1} & \textbf{1521.8} & \textbf{64.6} & \textbf{8.2} & \textbf{443.2} & \textbf{22.2} & \textbf{36.8} & \textbf{5,007.1} & \textbf{79.9} \\
\bottomrule
\end{tabular}

\label{tab:overall1-5}
\end{table*}

\begin{table*}[t]
\centering
\caption{Cost and Budget comparison between {\tech} and baselines across all tasks: Token in Millions.}
\setlength{\tabcolsep}{1pt}
\begin{tabular}{l|ccc|ccc|ccc|ccc|ccc}
\toprule
\multirow{2}{*}{} 
& \multicolumn{3}{c|}{\textbf{T1}} 
& \multicolumn{3}{c|}{\textbf{T2}} 
& \multicolumn{3}{c|}{\textbf{T3}} 
& \multicolumn{3}{c|}{\textbf{T4}} 
& \multicolumn{3}{c}{\textbf{T5}} \\
& \textit{Time} & \textit{Token} & \textit{Token/Alert}
& \textit{Time} & \textit{Token} & \textit{Token/Alert}
& \textit{Time} & \textit{Token} & \textit{Token/Alert}
& \textit{Time} & \textit{Token} & \textit{Token/Alert}
& \textit{Time} & \textit{Token} & \textit{Token/Alert} \\
\midrule
SWE 
& 0.1h & 0.12M & 0.03M
& 0.1h & 0.30M & 0.05M
& 0.1h & 0.28M & 0.08M
& 0.1h & 0.20M & 0.50M
& 0.1h & 0.21M & 0.19M \\
SWE+KB
& 0.1h & 0.17M & 0.04M
& 0.1h & 0.30M & 0.05M
& 0.1h & 0.35M & 0.09M
& 0.1h & 0.21M & 1.05M
& 0.1h & 0.36M & 0.26M \\ 
\makecell[l]{MAS+pipeline}
& 0.1h & 1.06M & 0.23M
& 0.1h & 1.54M & 0.26M
& 0.1h & 0.86M & 0.27M
& 0.1h & 1.22M & 1.09M
& 0.1h & 2.36M & 1.69M \\
\makecell[l]{MAS+orchestrator}
& 0.9h & 0.65M & 0.16M
& 0.9h & 1.88M & 0.19M
& 0.6h & 0.58M & 0.15M
& 0.4h & 0.32M & 0.19M
& 1.1h & 2.06M & 0.98M \\ \hline
\textbf{FirmHive} 
& \textbf{0.6h} & \textbf{2.00M} & \textbf{0.19M}
& \textbf{1.4h} & \textbf{5.66M} & \textbf{0.20M}
& \textbf{0.8h} & \textbf{5.03M} & \textbf{0.42M}
& \textbf{0.9h} & \textbf{1.55M} & \textbf{0.19M}
& \textbf{2.3h} & \textbf{17.43M} & \textbf{0.47M} \\
\bottomrule
\end{tabular}
\label{tab:overhead}
\end{table*}

Table~\ref{tab:sota_e2e_comparison} compares {\tech} with two SOTA firmware vulnerability tools.
{\tech} discovers a total of 1,802 alerts, significantly surpassing both \textsc{Mango} (1,109) and \textsc{SaTC} (644).
While \textsc{Mango} focuses primarily on control-impact (CI) and buffer overflow (BOF) vulnerabilities within binaries, {\tech} extends the analysis scope beyond binary code.
Specifically, among {\tech}’s total alerts, 1,206 correspond to CI/BOF issues, 192 are other binary vulnerabilities (e.g.,path traversal, unsafe memory operations), and 404 are non-binary alerts including XSS and authentication bypass vulnerabilities in web resources.
These additional categories are entirely missed by \textsc{Mango} and \textsc{SaTC}, which are limited to binary-level analysis.


To evaluate the reliability of {\tech}’s outputs in Task T5,
we manually reviewed a random sample of 300 alerts, including 200 labeled by {\tech} as true positives (TPs) and 100 labeled as false positives (FPs).
Table~\ref{tab:ai_verification} summarizes the manual validation results.
Among the 200 alerts automatically classified as TPs, 142 (71\%) were confirmed as genuinely exploitable vulnerabilities, while 42 (21\%) were false detections.
Conversely, among the 100 alerts classified as FPs, 26 (26\%) were found to be true vulnerabilities that {\tech} had conservatively dismissed.
These results yield a manual precision of 71\%, indicating that {\tech} achieves a high degree of reliability in autonomous vulnerability detection.
The 26 overlooked vulnerabilities suggest that although {\tech}’s internal validation mechanisms are largely effective, its conservative filtering criteria may occasionally exclude valid findings—an acceptable trade-off that favors precision over recall in large-scale firmware analysis.


\textbf{Summary.}
The evaluation highlights three major aspects in which {\tech} advances over existing firmware analysis systems.
(1) \textit{Paradigm.}  
{\tech} operates through an autonomous, LLM-driven workflow.  
This paradigm eliminates the need for manually defined pipelines and enables adaptive task reasoning during exploration.
(2) \textit{Capability.}  
{\tech} achieves cross-layer vulnerability discovery, analyzing binaries, scripts, and configuration files within a unified reasoning framework.  
(3) \textit{Performance.}  
{\tech} produces 1,802 alerts with an estimated precision of 71.0\%, reflecting both broader detection scope and higher confidence compared with traditional approaches.  
Detailed quantitative comparisons are provided in Appendix~\ref{appendix:toa_case} (Table~\ref{tab:paradigm_ultimate}).

\find{Finding 1: {\tech} presents a future direction for vulnerability discovery in firmware security analysis, belonging to an autonomous, AI-driven analysis paradigm rather than a traditional expert system.
}

\subsection{Comparison with Agent Baselines (RQ2)}
\label{sec:rq1}

For RQ2, we compare {\tech}'s performance against the SWE-Agent and MAS.  
All approaches use DeepSeek-v3.1 as the backend LLM.
We also use \textit{alert} to mean a potential security finding.
T1 alerts denote hard-coded secrets such as passwords or API keys;  
T2 alerts correspond to detected third-party components and their CVEs;  
T3 alerts involve tracing data flows from NVRAM or environment variables to security-critical operations;  
T4 alerts reflect web attack chains from untrusted input to dangerous sinks;  
T5 alerts represent vulnerabilities as RQ1 described.
Table~\ref{tab:overall1-5} summarizes the evaluation results across all tasks using three metrics per firmware: (1) average number of alerts, (2) reasoning steps, and (3) files accessed.

\textbf{Alert Coverage.}
{\tech} consistently identifies more alerts than all baselines across every task. 
For T1, it detects 10.3 alerts, $\thicksim$2.2× higher than baselines. 
For T2, 28.5 alerts, $\thicksim$4.6× higher than SWE+KB (6.2) and $\thicksim$3.1× higher than MAS+orchestrator (9.7). 
For T5, 36.8 alerts, $\thicksim$33–36× higher than SWE variants (1.1–1.4) and $\thicksim$17–36× higher than MAS baselines (1.4–2.1).

\begin{figure*}[!t]
    \begin{tabular}{ l   l  l }
    \begin{minipage}[t]{0.30\linewidth}
    \centering
    \includegraphics[width=0.9\linewidth]{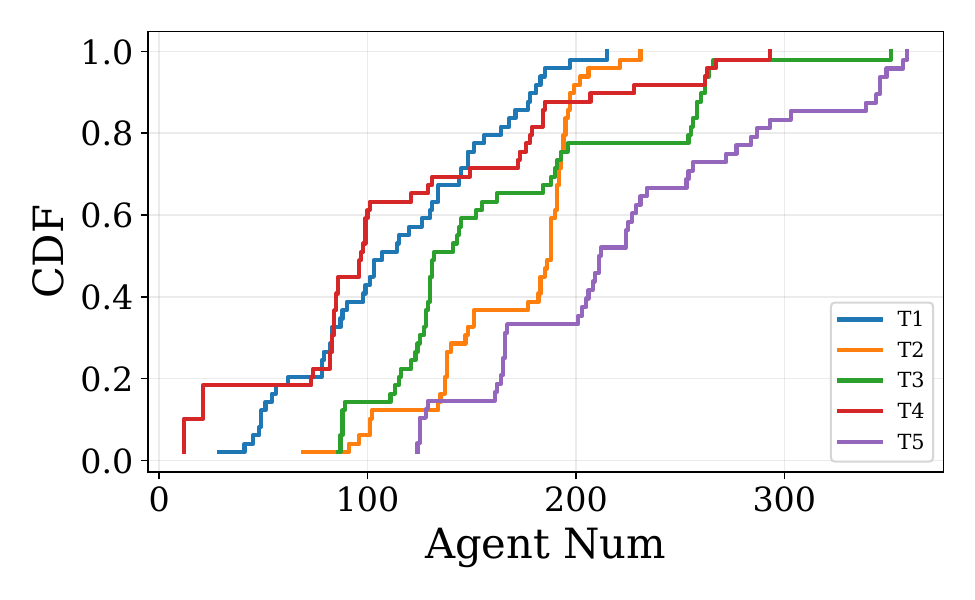}
    \caption{The CDF of the number of agents spawned in each task.}
    \label{fig:exp:agent-cdf}
    \vspace{-0.1in}
    \end{minipage} 
    &
    \begin{minipage}[t]{0.30\linewidth}
    \includegraphics[width=0.91\linewidth]{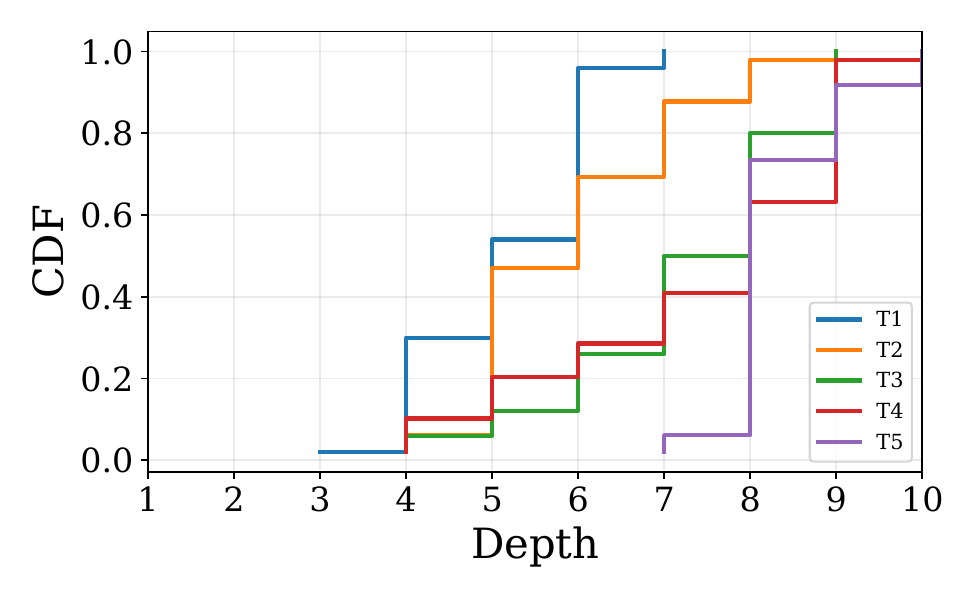}
    \caption{The CDF of the depth of the agent trees in each task.}
    \label{fig:exp:depth-cdf}
    \vspace{-0.1in}
    \end{minipage}
    & 
    \begin{minipage}[t]{0.30\linewidth}
    \includegraphics[width=0.91\linewidth]{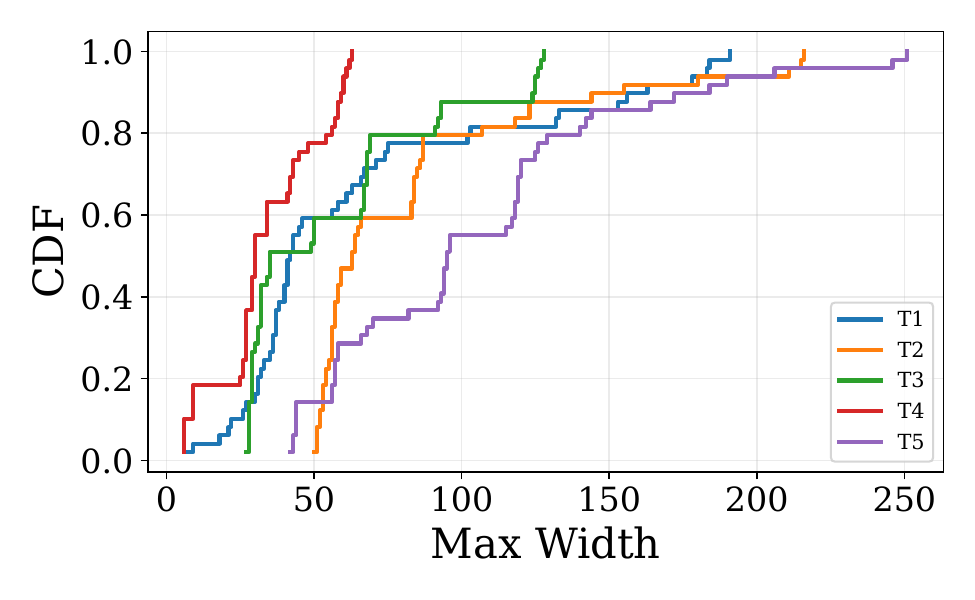}
    \caption{The CDF of the maximum branching factor in each task.}
    \label{fig:exp:max-branch-cdf}
    \end{minipage}
    \end{tabular}
\end{figure*}
\textbf{Reasoning Depth.} {\tech} performs substantially deeper analysis, with reasoning steps ranging from 692.5 (T1) to 5,007.1 (T5), representing $\thicksim$5.6–16× the depth of SWE variants (43–49) and $\thicksim$4–11× that of MAS baselines (124–430). 
The single-agent SWE exhibits shallow reasoning due to early termination, resulting in far fewer detected findings.

\textbf{Exploration Breadth.} {\tech} accesses 57–96.5 files for T1–T2 and 79.9 files for T5, $\thicksim$6–10× more than SWE (5–9) and comparable to MAS+pipeline. 
Unlike MAS+pipeline, which touches many files indiscriminately, {\tech} achieves proportional alert coverage per file. 
MAS+orchestrator reduces noisy exploration by centrally planning subtasks, yielding more meaningful analyses with fewer file accesses. Overall, MAS baselines improve over SWE in breadth and depth but remain far below {\tech}’s efficiency and effectiveness.


\textbf{Cost \& Budget}. 
Table~\ref{tab:overhead} summarizes the runtime and token usage across T1–T5.
{\tech} incurs substantially higher absolute compute and LLM usage than the single-agent baselines: per-firmware runtime and token consumption are both larger across all tasks (e.g., 2.3 h and 17.43 M tokens on T5).
When normalized by alert yield, {\tech}’s token-per-alert ratio remains moderate, ranging from 0.19 M to 0.47 M tokens across tasks.

Single-agent methods (SWE, SWE+KB) exhibit the lowest absolute cost and token consumption, but their minimal alert generation renders them inefficient in practice.
Among multi-agent approaches, MAS+pipeline often expends large token budgets without proportional alert yield due to noisy and shallow exploration.
MAS+orchestrator improves this by selectively allocating analysis steps toward targets.
Compared to these MAS baselines, {\tech} achieves the best overall \textit{token efficiency}.
Its token/alert (0.19–0.47 M) is roughly 3–4× lower than MAS+pipeline (0.23–1.69 M) and slightly better than MAS+orchestrator in most tasks.
In short, {\tech} incurs higher absolute runtime and token cost to achieve substantially greater discovery coverage, while maintaining competitive—and often superior—per-alert efficiency across diverse analysis tasks.

\textbf{Accuracy}. 
To validate the correctness of the alerts across all tasks, we performed a manual evaluation on a representative subset of firmware samples. 
Specifically, we randomly selected 50 alerts per task from the results of {\tech} and all baselines, totaling 250 alerts across the 5 tasks.
For each alert, we manually examined the raw filesystem and execution logic to confirm whether the reported alerts (e.g., hard-coded credentials, third-party components, NVRAM traces, or web attack flows) were accurate and complete.
{\tech} achieved the highest verification accuracy (82\%). 
By comparison, SWE‑Agent and SWE‑Agent + KB achieved 25\% and 26\%, respectively. 
MAS+orchestrator performed second best with an accuracy of 42\%, followed by MAS at 37\%.
These results highlight that {\tech}’s recursive delegation, structured ToA reasoning, and PKH-backed state tracking substantially improve reliability—especially in tasks.


\find{
Finding 2: 
Compared to existing agent architectures, {\tech} enhances the capabilities of LLMs by enabling them to handle complex, large-scale tasks more effectively.
}

\subsection{Firmware Analysis Stability (RQ3)}
\label{sec:rq2}

As we mentioned before, we leave the delegation decision and firmware analysis to the LLM, which introduces a level of unpredictability. 
For RQ2, we present workflow adaptability and quantitatively evaluate run‑to‑run stability.

\textbf{Workflow Adaptability.}
{\tech} can dynamically adapt workflows to different firmware and tasks.
We quantify this adaptability through three metrics: the total number of agents spawned (\textit{agent\_count}), the maximum ToA depth (\textit{depth}), and the maximum branching width (\textit{max\_branch}).

\textit{Agent Count.}
As shown in Figure~\ref{fig:exp:agent-cdf}, lighter tasks (T1–T4) typically spawn fewer than 150 agents, while complex firmware analysis tasks (T5) expand the ToA substantially, with a median of 248 and peaks above 350 agents.  
This demonstrates that {\tech} autonomously scales its reasoning capacity with task complexity instead of following a static pipeline.

\textit{Depth.}
Figure~\ref{fig:exp:depth-cdf} shows that the ToA depth remains stable (median 5–6) for simpler tasks but increases sharply for T5 (median 9, maximum 10).  
Deeper trees indicate that {\tech} engages in more prolonged, multi-stage reasoning to correlate evidence across layers of firmware components.

\textit{Branching Width.}
As illustrated in Figure~\ref{fig:exp:max-branch-cdf}, the maximum branching factor increases progressively across tasks—from medians of 19 (T4), 47 (T1), and 49–79 (T2–T3) in T5—reflecting broader parallel exploration at the same layer.  
Together, these results confirm that {\tech} adaptively expands both vertically (depth) and horizontally (breadth) to balance reasoning depth and coverage across diverse firmware structures.

Overall, {\tech} demonstrates pronounced structural adaptability: simpler tasks (T1–T3) yield compact ToAs with limited depth, while complex multi-stage reasoning tasks (T5) trigger extensive decomposition and inter-agent collaboration.
This dynamic topology formation enables {\tech} to autonomously balance reasoning granularity and exploration breadth at runtime.
In contrast, static and pre-defined pipelines such as MAS lack the flexibility to adjust to varying firmware complexity or task composition.
In firmware analysis, such adaptability is critical—since the number of relevant components, dependencies, and call-chain depth are often unknown a priori.
We further visualize 3 ToAs to illustrate {\tech}’s adaptive topology formation under different task complexities (see Appendix~\ref{appendix:toa_case}).

\textbf{Stability of {\tech}.}
A common concern for LLM-driven systems is stochastic reasoning, which may lead to inconsistent agent behaviors and divergent workflows across runs.
To assess this effect, we evaluate {\tech} on 10 representative firmware images using three backbone LLMs (DeepSeek-V3.1, Claude-3.7-Sonnet, GPT-4o, and Gemini-2.5-Flash).
Table~\ref{tab:stability_summary} summarizes the results.
Across all tasks, {\tech} consistently meets the predefined stability thresholds: both the number of spawned agents and the number of generated alerts exhibit minimal variance.
Furthermore, the Jaccard similarity between alerts from independent runs remains high—ranging from 0.84 (T1) to 0.76 (T5)—demonstrating that the system repeatedly converges to near-identical ToA structures and reproduces similar findings.

Although delegation is determined dynamically by LLMs at runtime, these decisions are not unconstrained.
{\tech} couples each reasoning step with a fixed context and consistent environmental observations, effectively bounding the LLM's delegation.
These results show that as long as the underlying LLM maintains sufficient reasoning competence, {\tech} achieves stable delegation behavior and reproducible analytical outcomes across repeated executions.


\begin{table}[!t]
\centering
\small
\caption{Stability summary in the same firmware subset: 10 different runs and 3 different LLMs.}
\label{tab:stability_summary}
\begin{tabular}{lcc}
\toprule
Task & avg. number of agents $\pm$ std  & alerts $\pm$ std  \\
\midrule
T1 & 55.70 $\pm$ 3.80  & 10.2 $\pm$ 3.10 \\
T2 & 84.22 $\pm$ 5.10    & 28.5 $\pm$ 2.80 \\
T3 & 67.18 $\pm$ 4.60 & 12.1 $\pm$ 1.40 \\
T4 & 40.32 $\pm$ 2.50    & 8.2 $\pm$ 0.40 \\
T5 & 189.7 $\pm$ 12.50  & 36.8 $\pm$ 4.90 \\
\bottomrule
\end{tabular}
\end{table}

\find{Finding 3: The system's primary advantage is that it adapts to different tasks and firmware images, and its behavior is stable rather than arbitrary.}

\subsection{Component Contribution Analysis (RQ4)}
\label{sec:rq3}


\textbf{LLM Selection}.
We evaluate the following LLMs as the backbone for {\tech}: GPT-4o, DeepSeek v3, Claude-3.7-sonnet, and Gemini-2.5-flash on a representative subset of 10 firmware images. All models are accessed through their official APIs and integrated into {\tech}. 
Specifically, we use the task T5 to evaluate the performance of these LLMs in the context of firmware analysis. 
Table~\ref{tab:llm_verification_eval_revised} summarizes the performance of these LLMs for finding vulnerabilities in firmware.
Overall, our findings suggest that as long as the selected LLM possesses sufficient single-step reasoning capability, the performance of {\tech} remains largely robust to the specific choice of backbone model.

\begin{table}[!t]
\small
\centering
\caption{{\tech}'s Performance with different backend LLMs.}
\label{tab:llm_verification_eval_revised}
\setlength{\tabcolsep}{5pt}
\begin{tabular}{lcc}
\toprule
LLM Model & \begin{tabular}[c]{@{}c@{}}Alert  Count\end{tabular} & \begin{tabular}[c]{@{}c@{}} Precision (\%)\end{tabular} \\
\midrule
GPT-4o & 385 & 68.5 \\
Claude-3.7-Sonnet & 312 & 73.0 \\
DeepSeek-V3.1 & 333 & 70.1 \\
Gemini-2.5-Flash & 345 & 69.1 \\
\bottomrule
\end{tabular}
\end{table}

\begin{figure}[!t]
    \centering
    \includegraphics[width=0.75\linewidth]{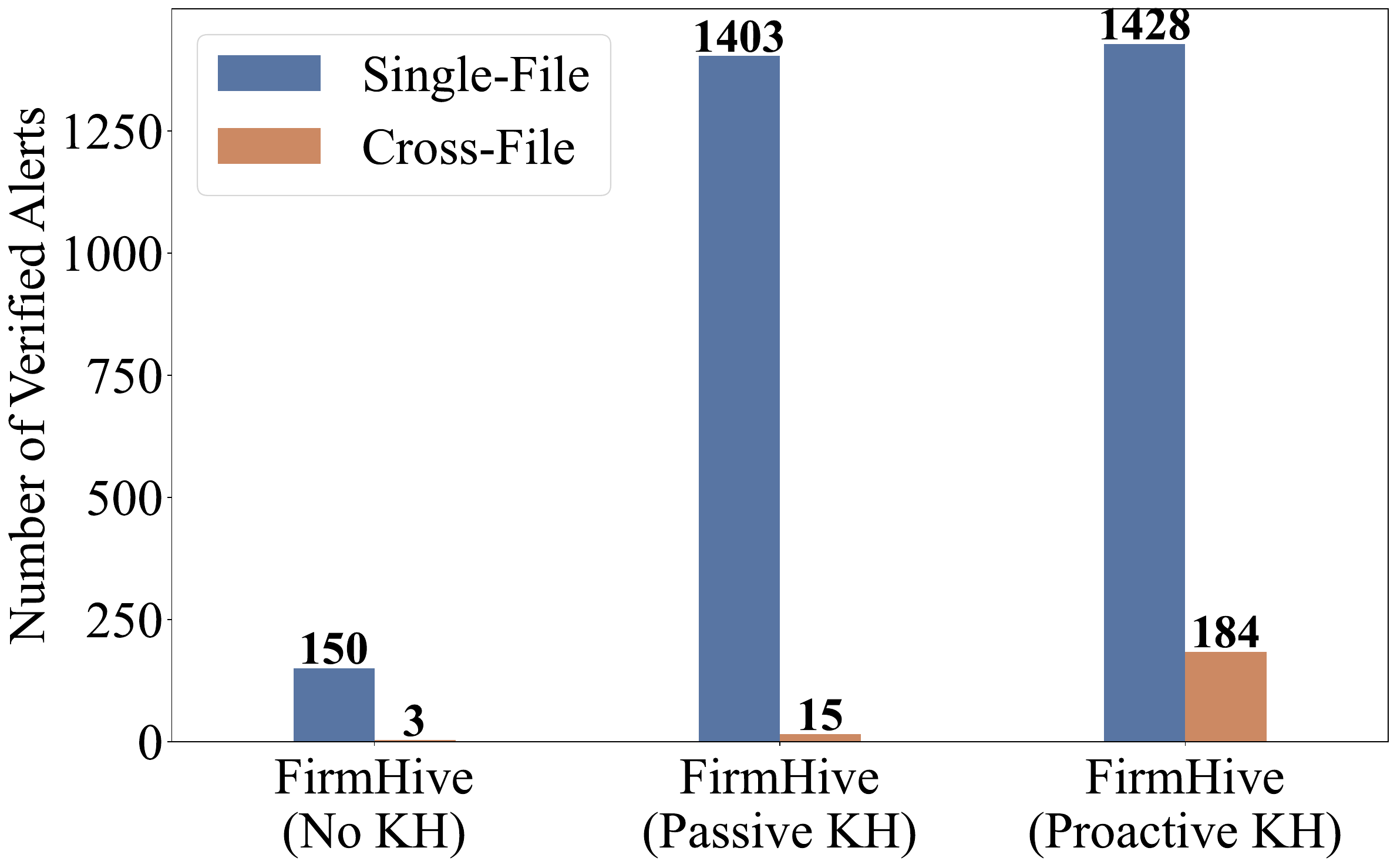}
    \caption{Performance comparison of {\tech} on Task T5 with and without the knowledge agent.}
    \label{fig:kb:perf}
\end{figure}

\textbf{PKH \& KB Ablation}.
We first assess the impact of a passive KB on reasoning performance.
As shown in Table~\ref{tab:overall1-5}, SWE-Agent and SWE+KB exhibit nearly identical reasoning depth and coverage across all tasks, indicating that storing past findings alone does not enhance multi-step reasoning.
Instead, KB can slightly improve SWE-Agent's alert count (e.g., from 4.2 to 4.4 in T1, from 1.1 to 1.4 in T5) because it helps the agent to record and recall prior findings.

\begin{table}[!t]
\centering
\small
\caption{Impact of dynamic delegation.}
\label{tab:ablation_delegation}
\begin{tabular}{lcccc}
\toprule
 &   Alert & Agent\_count  &  Jaccard   \\
\midrule
MAS + pipeline & 0.40 $\pm$ 0.20 & 1.0 $\pm$ 0.0  & 0.11 \\
MAS + orchestrator     & 3.30 $\pm$ 1.10 & 150.0 $\pm$ 20.0 & 0.16 \\
{\tech}        & 36.8 $\pm$ 4.90 & 189.7 $\pm$ 12.5 & 0.78 \\
\bottomrule
\end{tabular}
\end{table}

We further conduct an ablation study comparing three variants of {\tech}:
(1) without KB, 
(2) with a passive KB, and 
(3) with the PKH. 
As shown in Figure~\ref{fig:kb:perf}, PKH boosts {\tech}’s performance on T5 (vulnerability discovery), especially for multi-file vulnerabilities.
While passive storage yields comparable results on single-file cases, PKH’s active reasoning mechanism enables cross-file correlation and cumulative evidence synthesis, resulting in substantially higher detection accuracy.

\textbf{Impact of Dynamic Delegation}. 
We compare three configurations on a representative subset of 10 firmwares (Task T5), fixing DeepSeekV3 as the backbone and all budgets (S\_max=30, max\_depth=4). 
Each (firmware, configuration) pair is repeated R=10 times.  
Measured metrics: verified alerts, agent\_count, median pairwise Jaccard of verified alerts.
Table~\ref{tab:ablation_delegation} summarizes the results.
Dynamic delegation yields a large increase in verified findings while maintaining equal or lower relative variability and substantially higher pairwise Jaccard, indicating that the extra exploration is systematic (input‑driven) rather than arbitrary. 
Static workflows and single‑agent runs produce far fewer verified alerts and lower overlap, confirming delegation as the primary contributor to coverage gains.


\section{Issues and Lessons Learned}
\label{sec:discuss}

\textbf{Issue 1: Dataset Validity}.
A potential threat to validity lies in whether the dataset used in our evaluation adequately represents the diversity and complexity of the broader firmware ecosystem. 
On one hand, the dataset with labeled vulnerabilities has been widely used in prior works, including Karonte~\cite{redini2020karonte}, Mango~\cite{gibbs2024operation}, and SaTC~\cite{chen2021sharing}. On the other hand, we acknowledge that the dataset may still underrepresent rare architectures, proprietary firmware formats, or the latest devices.
To address this issue, we will continuously supplement our experiments with additional firmware from public sources in the future.

\textbf{Issue 2: LLM Hallucination}.
LLMs are known to occasionally generate hallucinated content, which is syntactically plausible but incorrect outputs.  
Note that {\tech} focuses on extracting and reasoning over existing firmware artifacts rather than generating new content, thereby reducing the surface for hallucination.
To further lower this risk, every claimed finding must be linked to deterministic evidence (e.g., tool outputs such as radare2 or grep, or entries in the PKH).
These practical measures substantially reduce the chance of hallucination, though they cannot entirely eliminate it.


\textbf{Issue 3: Coverage}.
Information omissions are inevitable for both traditional tool-based methods and LLM-based approaches. 
The core reason is that firmware contains a vast amount of diverse and complex information, and no single method can cover everything. 
Our system may miss information for three main reasons: (1) some firmware files or components are not targeted, so agents never analyze them deeply; (2) some analysis tasks exceed the current LLM’s capabilities, e.g., complex cross-function dataflow, or hidden call chains; (3) our tools have blind spots and cannot reveal information for firmware artifacts.


\textbf{Lesson 1.} 
Contrary to intuition, increasing the number of agents does not inherently lead to higher token consumption or longer runtime. 
In our experience, token and time costs scale with the intrinsic complexity of the analysis task, including firmware size, inter-component coupling, and the depth of reasoning required.
The choice between a single agent and multiple agents does not change these fundamental factors. 
Deploying multiple agents is primarily an architectural strategy for distributing and parallelizing reasoning.

\textbf{Lesson 2.}
Intermediate results are important for security tasks; they include temporary outputs, context snippets, transient variables, and other intermediate artifacts. 
Handling these intermediate results needs to be dynamic and adaptive as analysis progresses and new information emerges.
{\tech} enables the LLM to access and use these intermediate signals dynamically. This is the main distinction from traditional pipelines or existing LLM-based approaches, which filter intermediate results using static rules or code.
Notably, {\tech} contains no handcrafted analysis code or logic, yet it achieves better analysis performance, underscoring the importance of intermediate results.

\textbf{Lesson 3.}
Maintaining the LLM’s capability over time is critical for applying LLMs to security tasks.
Current LLMs often outperform traditional methods on short tasks (e.g., a code snippet, a single file, or a single function). However, security analysis involves a large amount of code and many files, and the process becomes long and complex. 
Thus, the LLM’s effectiveness is hard to maintain, and it may degrade over time (context rot). 

\section{Related Work}
\label{sec:related}

\subsection{LLM Agents}

\textbf{Large Language Models (LLMs)} based on Transformer architectures (e.g., GPT-4~\cite{openai}, Claude~\cite{claude}, LLaMA~\cite{llama}, Codex~\cite{open-codex}) have demonstrated strong reasoning, planning, and generalization capabilities. Their performance depends heavily on how prompts are crafted. Advanced reasoning methods include task decomposition~\cite{wang2022self}, chain-of-thought prompting~\cite{wei2022chain}, and decision trees via Tree-of-Thought~\cite{yao2023tree}. Techniques like Reflexion~\cite{shinn2023reflexion} and Chain of Hindsight~\cite{liu2023languages} introduce self-feedback to improve output quality.

\textbf{LLM-based Agents} use LLMs as a central controller to perceive environments, decompose tasks, and take actions through external tools~\cite{park2023generative, wang2024survey, xi2023rise}. Key modules include: (1) \textit{planners} that create task graphs~\cite{wang2023voyager}; (2) \textit{perception modules} enabling multi-modal input~\cite{yao2022react}; and (3) \textit{action modules} that expand capabilities via tool usage~\cite{huang2022large, schick2023toolformer, lu2024chameleon}.
LLMs have shown strong potential in software engineering (e.g., GitHub Copilot~\cite{copilot}) and security tasks~\cite{chen2023teaching, feng2023prompting, pei2023can, pearce2022pop}. For instance, Feng and Chen~\cite{feng2023prompting} used LLMs to replay Android bugs, while others have explored vulnerability discovery and reverse engineering~\cite{li2023hitchhiker, pearce2023examining}.

\subsection{Embedded Firmware}

\textbf{Security Analysis}. Firmware exposes a broad range of attack vectors: (1) \textit{third-party libraries} with known CVEs~\cite{zhao2023uvscan, zhao2022large}; (2) \textit{hardcoded secrets} like passwords and keys~\cite{costin2014large}; (3) \textit{config files and certificates} that can expose services~\cite{firmwalker}; (4) \textit{open services} such as HTTP or FTP that attackers can exploit~\cite{costin2016automated}.
Chen et al.~\cite{chen2021sharing} proposed a novel input-centric strategy that identifies bugs in embedded systems by extracting and sharing common input keywords across different devices, reducing reliance on heavyweight static analysis.

\textbf{Vulnerability Discovery}. Prior work uses static and dynamic analysis to detect firmware flaws~\cite{feng2016scalable, chen2016towards, redini2020karonte}. 
Mango~\cite{gibbs2024operation} introduced a scalable taint-style analysis pipeline to identify vulnerabilities in binary firmware services at scale. Redini et al. \cite{redini2020karonte} explored inter-binary taint propagation, while others applied symbolic execution and modular program analysis \cite{feng2020p2im, yang2022modx, helmke2023towards}.
Prior studies lack the adaptive planning and deeper semantic understanding offered by LLM-based agents. Our work advances this direction by enabling dynamic, task-driven analysis.

\section{Conclusion}
\label{sec:conclusion}

In this paper, we presented {\tech} to address the inherent complexities of firmware security analysis. 
Unlike prior static and manual approaches, {\tech} leverages recursive agent generation, a multi-level orchestration, and a centralized knowledge base to overcome critical challenges such as long-term reasoning, large-scale exploration, and cross-component dependency resolution. 
Through extensive real-world experiments on diverse firmware analysis tasks, {\tech} demonstrated superior performance over existing LLM-based agents and traditional security tools, autonomously issuing about 1,802 high-confidence alerts with notable precision.
Our results highlight that LLM agents can transform cybersecurity tasks from a labor-intensive and error-prone process into an automated workflow.



\small
\bibliographystyle{plain}
\bibliography{ref/ref_online, ref/ref_security, ref/ref_firmware, ref/ref_llm, ref/ref_measurement}



\appendix

\subsection{System Implementation}
\label{appendix:imple}

\textbf{Implementation.}
We implemented a prototype of {\tech} in Python to demonstrate its feasibility. 
The system consists of five key components:

(i) LLM Wrapper.
A unified client layer that abstracts interactions with backend LLMs. It manages connection states, retry and backoff mechanisms, token accounting, and logging. So far, {\tech}'s LLM wrapper is compatible with several mainstream LLM providers, including: OpenRouter, Hugging Face Route, Azure, OpenAI, and Ollama.

(ii) Response Parsing.
{\tech} enforces structured outputs by extracting JSON objects from raw LLM responses using regex parsing, schema validation, and a bounded feedback–retry loop to ensure format robustness and compliance.

(iii) Memory Storage.
Agent–LLM interaction histories are stored as lists of $(\texttt{role}, \texttt{content})$ pairs. Roles include \textit{system}, \textit{user}, \textit{assistant}, \textit{tool}, and \textit{error}, enabling contextual continuity and error tracing across sessions.

\begin{table}[t]
\small
\centering
\caption{Usage tools for firmware analysis.}
\label{tab:tool}
\begin{tabular}{p{0.22\linewidth}p{0.68\linewidth}}
 \toprule
  \textbf{Tool} & \textbf{Description} \\
  \midrule
  GetContextTool & Fetches context (e.g., from a specific root path) to build a "sub-sandbox" view. \\
  ShellExecuteTool & Executes non-invasive shell commands (e.g., `ls`, `file`, `grep`) \\
  Radare2Tool & Wraps core Radare2 commands for static binary analysis, including `pdg` and all other commands. \\
    \midrule
 DelegationTool  & Generate a subtask, Spawns a child agent to handle a subtask, and Runs a child agent (1$\rightarrow$1). \\
 Parallel DelegationTool  & Generate a list of subtasks,
 Spawns multiple child agents in parallel, and Runs children agents (1$\rightarrow$n). \\ \hline
 StoreTool  & Stores validated findings in the Persistent Knowledge Hub (PKH). \\
    QueryTool  & Queries the PKH for specific information. \\
    ExploreTool  & Enumerates possible findings that may exist potential correlations with the current context.\\
 \bottomrule
\end{tabular}
\end{table}

(iv) Agent Prompts and Tools.
{\tech} defines three role-specific system prompts—\textit{directory}, \textit{file}, and \textit{function}—to guide agent behavior. Agents operate via structured tool calls following a unified JSON schema
$(\mathtt{thought}, \mathtt{action}, \mathtt{action\_input}, \mathtt{status})$,
where $\mathtt{thought}$ records internal reasoning, $\mathtt{action}$ specifies the invoked tool, $\mathtt{action\_input}$ provides parameters, and $\mathtt{status}$ captures execution success or failure.
All available tools are summarized in Table~\ref{tab:tool}.

(v) Blueprint Execution.
The blueprint defines a script-based pipeline organizing agents into a three-layer delegation hierarchy. It instantiates a root agent that accepts user tasks and dynamically constructs a ToA at runtime.

Unlike existing frameworks such as LangChain, LangGraph, or CrewAI, {\tech} is implemented entirely from scratch. 
Firmware analysis requires fine-grained control over agent orchestration and dynamic delegation, which are difficult to achieve through existing frameworks.

\subsection{Supplementary Experiments}
\label{appendix:toa_case}

\begin{table}[!t] \small
  \makeatletter
  \centering
  \caption{The evaluation dataset. Each row includes the vendor/product, firmware count, directory count, and file count.}
  \label{tab:dataset}
  \begin{tabular}{l l r r r }
    \toprule
    Vendor (Product) & \# Firmware & \# Dir & \# Files \\
    \midrule
    NETGEAR (R/XR/WNR)  & 17   & 2,406  & 34,603 \\
    D-Link  (DIR/DWR/DCS) & 9    & 1,175  & 16,383 \\
    TP-Link (TD/WA/WR/TX/KC)  & 16   & 1,154  & 9,524 \\
    Tenda  (AC/WH/FH) & 7   & 459  & 3,394 \\
    \bottomrule
  \end{tabular}
\end{table}

\begin{table}[!t]
\small
\centering
\renewcommand{\arraystretch}{1.25}
\caption{Paradigm, Capability, and Performance Comparison between {\tech} and \textsc{Mango}.}
\label{tab:paradigm_ultimate}
\begin{tabular}{p{0.2\linewidth} p{0.25\linewidth} p{0.4\linewidth}}
\toprule
Paradigm & \textbf{\textsc{Mango}} & \textbf{{\tech}}\\
\midrule
Usage Model & Expert-Driven & \textbf{LLM-Driven} \\ 
Workflow & Fixed Pipeline & \textbf{Autonomous} \\
Intermediate results & Not Accessible & \textbf{Observable} \\ 
\midrule
\multicolumn{3}{l}{\textit{Capability}} \\
\midrule
Extensibility & T5 Only & \textbf{T1–T5 (Generalizable)} \\ 
Vulnerability & Binary (CI/BOF) & \textbf{Binary + Non-Binary (XSS, Auth Bypass, etc.)} \\ 
\midrule
\multicolumn{3}{l}{\textit{Performance}} \\
\midrule
Alerts & 1,109 & \textbf{1,802} \\
Precision & 47.9\% & \textbf{71.0\%} \\
\bottomrule
\end{tabular}
\end{table}

\begin{figure}[t]
    \centering
    \includegraphics[width=\linewidth]{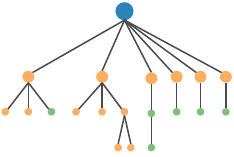}
    \caption{ToA structure for Archer\_C2\_V1\_170228 in Task T1.}
    \label{fig:topo1}
\end{figure}

\begin{figure}[t]
    \centering
    \includegraphics[width=\linewidth]{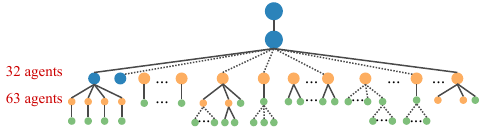}
    \caption{ToA structure for Archer\_C2\_V1\_170228 in Task T2.}
    \label{fig:topo2}
\end{figure}

\begin{figure}[t]
    \centering
    \includegraphics[width=\linewidth]{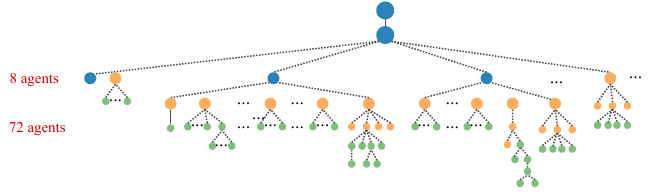}
    \caption{ToA structure for Archer\_C20\_V1\_151120 in Task T5.}
    \label{fig:topo3}
\end{figure}

To provide an intuitive understanding of {\tech}’s adaptive orchestration behavior, we visualize the Tree of Agents (ToA) generated under representative tasks of varying complexity.
Figures~\ref{fig:topo1}–\ref{fig:topo3} show the ToAs formed for (i) Task T1 and T2 using the same firmware (\textit{Archer\_C2\_V1\_170228}), and (ii) Task T5 using a more complex firmware (\textit{Archer\_C20\_V1\_151120}).
Each node denotes an instantiated agent, and directed edges represent delegation relationships.
The ToA structure directly reflects how {\tech} decomposes the original task and coordinates its subtasks at runtime.

In T1, {\tech} constructs a compact ToA (22 agents, depth 4, width 11), where the root agent handles basic enumeration and a small number of sub-agents are spawned for contextual analysis.
In T2, the reasoning becomes more exploratory: the ToA grows to 139 agents with increased branching (depth 5, width 63), corresponding to broader code coverage and multi-component correlation.
For the complex vulnerability discovery task T5, {\tech} dynamically expands its reasoning scope, producing 150 agents with a deeper hierarchical structure (depth 9, width 72).
This deeper, wider ToA captures recursive exploration and multi-stage reasoning across binaries, scripts, and configuration files.

These qualitative results align with the statistical observations in Figure~\ref{fig:exp:agent-cdf}–\ref{fig:exp:max-branch-cdf}, confirming that {\tech}’s topology is not pre-defined but self-constructed at runtime based on firmware complexity and task semantics.
Such dynamic topology formation is essential for firmware analysis, where neither the number of relevant components nor their dependency depth can be determined a priori.

\subsection{System Prompt}
\label{prompt}

This appendix lists the exact system prompts used for the Directory, File, and Function agents.
These prompts (List~\ref{lst:directory-prompt}, \ref{lst:file-prompt}, \ref{lst:cca-prompt}) are fixed throughout all experiments to ensure reproducibility and stable delegation behavior.

\begin{table}[ht] \footnotesize
  \centering
  \caption{Directory Agent's system prompt.}
  \label{lst:directory-prompt}
\begin{tabular}{p{8.1cm}}
\toprule
\color{blue}{system: $\|$} \\   
You are a firmware filesystem static analysis agent. 
Your task is to explore and analyze based on the current analysis focus (a specific directory). 
Please stay focused on the current target; when you believe the analysis is complete or no further progress can be made, continue to the next task or end the task.

\color{blue}{Working principles:} \\

1.  **Understand the Requirements**\\
 *   Always focus on the specific task while referring to the user's overall or initial goal. 
        Note: if the task does not match the current analysis focus, you must stop the analysis and promptly report to the user. 
        Do not perform cross-directory analysis.\\
    *   Carefully understand what firmware content and objectives the user wants to analyze. 
        Unless you are certain they do not meet the user’s requirements, do not skip directories or files that might be relevant. 
        If the user’s intent is unclear, choose the best analysis path based on firmware characteristics, decompose complex tasks, and reasonably invoke analysis assistants. \\
    *   If the user’s intent is unclear, select the optimal analysis path according to firmware characteristics, decompose complex tasks appropriately, and invoke analysis assistants when needed.

2.  **Formulate an Analysis Plan**\\
    *   Choose the optimal analysis path based on the characteristics of the firmware. \\
    *   For complex tasks, decompose them into multiple subtasks with clear objectives and sub-targets, and appropriately invoke assistants and tools. When deeper analysis is required, invoke suitable assistants: \\
        *   **Explore Directory**: use the subdirectory analysis assistant or its parallel version to switch to the specified directory for analysis. \\
        *   **Analyze File**: use the file analysis assistant or its parallel version to analyze the specified file. \\
    *   Adjust the analysis plan according to assistant feedback to ensure completeness and accuracy. 
        If the assistant cannot complete the task, re-formulate the plan. 
        If the assistant still fails after two attempts, move on to the next task. 

3.  **Problem Handling During Analysis**\\
    *   Record any technical difficulties and unique challenges encountered during analysis. \\
    *   Assess the impact of these issues in a real firmware environment. \\
    *   Use certain tools with caution to avoid excessively long outputs that may cause analysis failure, e.g., the `strings` tool. 

4.  **Submitting Analysis Results**\\
    *   Summarize all analysis results and respond to the questions corresponding to the current task. \\
    *   Honestly report cases where evidence is insufficient or uncertain (specify what evidence or information is missing).\\

\bottomrule
\end{tabular}
\end{table}

\begin{table}[ht] \footnotesize
  \centering
  \caption{File Agent's system prompt.}
  \label{lst:file-prompt}
\begin{tabular}{p{8.1cm}}
\toprule
{\color{blue}{system: $\|$}}

You are a specialized file analysis agent. Your task is to perform in-depth analysis of the currently specified file and provide detailed, evidence-supported analysis results. Please focus on the current focus file or the current specific task. When you believe the analysis is complete or no further progress can be made, proceed to the next task or end the task.

\color{blue}{Working principles:} \\
- **Evidence-based:** All analyses must be grounded in actual evidence obtained from tools; speculative conclusions without evidence are strictly prohibited. \\
- **Result validation:** Critically evaluate the results returned by delegated subtasks (such as function analysis), and always verify their authenticity and reliability to prevent false results from contaminating the final conclusion.

\color{blue}{**Workflow:**}\\
1. **Understand the task:** Focus on the specific task of the current analysis file, while fully referencing the user's overall requirements. Note: if the task does not match the current analysis focus, you must stop the analysis and promptly report to the user. Do not perform cross-directory analysis.\\
2. **Perform analysis:** Ensure your analysis has sufficient depth. For complex tasks, decompose them into multiple subtasks with clear objectives and steps, and reasonably invoke analysis assistants or tools in sequence or in parallel. Choose the most appropriate tools or assistants to obtain evidence. Use certain tools cautiously (such as the `strings` tool) to avoid overly long results that may cause analysis failure. \\
3. **Completion and reporting:** After completing the analysis, use the `finish` operation and strictly follow the format below to submit your final report.

\color{blue}{**Final response requirements**:} \\
* Answer all questions related to the current task, and your response must be fully supported by evidence. Do not omit any valid information.\\
* Support all findings with concrete evidence, and truthfully report any insufficiencies or difficulties in obtaining evidence. 
{\color{blue}{\{FILE\_RESPONSE\_FORMAT\_BLOCK\}}} \\
* In the 'action' field, select a tool or 'finish', and provide the corresponding parameters or final response in the 'action\_input' field.\\

\bottomrule
\end{tabular}
\end{table}

\begin{table}[ht] \footnotesize
  \centering
  \caption{Function Agent's system prompt.}
  \label{lst:cca-prompt}
\begin{tabular}{p{8.1cm}}
\toprule
{\color{blue}{system: $\|$}}

You are a dedicated function call chain analysis assistant. Your core responsibility is to track the flow path of the tainted data within the current function. Please focus on the current function, and when you think the analysis is complete or no further progress can be made, continue to analyze the next task or end the task.

\color{blue}{*Working principles*:} \\

- All analysis is based on actual evidence, without unfounded guesses. \\
- When a problem is found, clearly indicate its location, trigger conditions, and potential impact.

\color{blue}{*Analysis process*:} \\

1. **Focus on the current function**: \\
* Analyze the function code, understand its parameters, return values, and how it handles the incoming tainted data. \\

2. **Taint flow judgment and decision**: \\
* **Does the taint flow into the sub-function?** \\
* **Yes**: Create a delegate task for the sub-function. In the task description, it must be clearly stated: \\
1. **Target function**: The address of the sub-function to be analyzed. \\
2. **Taint parameter**: Which parameter is tainted. \\
3. **Taint transformation**: What processing or transformation the tainted data has undergone before passing it. \\
* **No**: If the tainted data is eliminated in the current function or not passed to the child function, terminate the analysis of this path. \\

3. **Analysis Depth Control**: \\
* When the tainted data has been fully verified and purified (such as strict boundary checks), you can stop tracing. \\
* For known safe library function calls or complex situations that cannot be analyzed, stop going deep and report the current situation.

4. **Summarize the results**: \\
* After all path analysis is completed, summarize the final results, including the complete taint propagation path, discovered vulnerabilities, and use `finish` to end the task. 
{\color{blue}{\{FUNCTION\_RESPONSE\_FORMAT\_BLOCK\}}}\\

\bottomrule
\end{tabular}
\end{table}

\end{document}